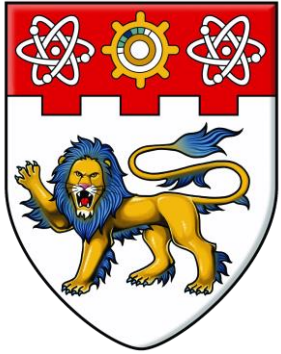

**Semi-automatic conversion from OSG to CityGML**

**Pranjal Swarup**

**SCHOOL OF COMPUTER SCIENCE AND ENGINEERING**

**2016**

# Semi-automatic conversion from OSG to CityGML

Pranjal Swarup

(G1502107H)

**SCHOOL OF COMPUTER SCIENCE AND ENGINEERING**

A Dissertation Submitted in partial fulfilment of the
requirements for the degree of    Master of Science in
Digital Media Technology

Supervised by
Associate Professor Jianmin Zheng
2016

# TABLE OF CONTENTS

# TABLE OF FIGURES

# Abstract

CityGML is a data model used to represent the geometric and semantic information of urban 3D city objects. There are several ways to generate 3D models for applications such as gaming, media content for movies and TV and 3D printing among others. Since visualization is the primary purpose of these methods of 3D model generation the lack the sematic information required for applications such as spatial data-mining, thematic queries and geospatial simulation and analysis. Therefore, there is a need to develop methods for mapping of sematic information of new and existing 3D geometric models and their conversation to CityGML formats so it can be stored for future use.

The conversion of must allow mapping of classes of objects with a higher Level-Of-Detail (LOD) such as LOD3 and LOD4 which represent an urban city model in greater details. A methodology of the conversion is developed and a prototype is implemented to validate the algorithms and the methods developed for this project. The methods can be broadly categorized into three parts a) Extraction of geometric information; b) Segmentation of the model into various groups for semantic mapping; c) Mapping of semantic information on the model. Topological aspects are considered for the mesh segmentation to allow users to easily select and map semantic information to the geometric model. The generated models can be used in a wide range of applications ranging from disaster management to urban simulations such as rooftop solar panel installations.

# CHAPTER 1

# INTRODUCTION

## BACKGROUND

3D architectural models are crucial to virtual reality applications. They have a multitude of applications such as visualization of an environment, running simulations such as energy demand estimation and urban planning such as examining the solar installation potential on rooftops[1].

Architectural models can be derived using several different methods including reconstruction of point-cloud obtained from drone mounted laser cameras, stereoscopic cameras, extrusion from satellite imagery, procedural and polygonal modelling. These meshes can be stored in a 3D graphics file format of choice such as DAE(Collada), OBJ(Wavefront) or OSG(OpenSceneGraph), among others. However while these 3D graphics file formats contain the geometrical and appearance properties, they do not have any semantic information. The conversion to CityGML requires geometric transformations and semantic mapping that can be output to a syntactically correct XML representation valid according to the CityGML standard.

# SCOPE OF THE PROJECT

The aim of the project is to:

- Conduct a feasibility study on conversion of 3D mesh representation in OpenSceneGraph/OBJ format to the CityGML format.

- Develop a method for semi-automatic semantic mapping of a polygonal model, apply geometric transformations, tag the polygons with its corresponding label, and output XML representation of the converted CityGML LOD3 model.

- Segmentation techniques such as shape analysis, segmentation using normal vectors are to be used to minimize the user input required for the sematic mapping of the model being converted.

The methods developed during the project are aimed at geometrically valid and semantically meaningful conversion of polygonal mesh from OpenSceneGraph format to CityGML LOD3 format. While user input is required to for mapping of the model, the project aims to provide methods and tools to select the semantically correlated surfaces.

While CityGML supports several classes of city models such as transportation systems, terrain and water bodies etc., most of these classes of objects are out of the focus of this project. The project focuses on the building model and the user input-driven conversion of its exterior shell geometry into CityGML LOD3 model. Appearance properties such as textures and materials have not been covered in this project, the solid geometry that fill up the LOD3 model interiors have been omitted as this project focuses on the multi-surface meshes such that it can be extended to LOD4 models that incorporates the internal surfaces such as the interior wall surfaces.

A gml:MultiSurface representation of LOD 3 building requires that there are no opening in the exterior shell, however in this project this is not validated and it is assumed that the mesh does not containing any opening or they are closed using ClosureSurface polygons, which can be semantically mapped.

# ORGANIZATION OF THE DISSERTATION

Background information on CityGML, OpenSceneGraph and format conversion to CityGML format is given in Chapter 2. Chapter 3 covers the conversion methodology including geometric transformations, pre-processing and segmentation. Chapter 4 discusses the implementation of the prototype, tools provided for easy segmentation and mapping of semantic data, results from the prototype and limitations of the approach. Finally in Chapter 5 a summary of the project is presented along with the recommendations for future work and research on semi-automatic conversion of geometric models to CityGML format.

# CHAPTER 2

# RELATED WORKS

## CityGML

### General characteristics of CityGML

Construction of virtual building models is a recent phenomenon that is mainly used for urban information modelling purposes. Civic bodies such as municipalities, organizations such as large companies and universities require virtual 3D models for various purposes including energy resources allocation, navigation planning, simulation of natural environment, business and tourism development, radio network planning and facilities management. Several methods such as aerial imagery, CAD modelling, laser camera etc. are used to register the geometric information to create three-dimensional model of a city, however only after adding semantic attribute to the 3D data can the model be stored as a CityGML object. CityGML objects help run simulation that are not only dependent on geometrical models but need semantic information to run context based simulations, running spatial data-mining and analysis.

CityGML defined as an application schema of Geography Markup Language(GML), which is an OGC standard markup language based on XML. GML attributes are used for geometric data representations in CityGML.

The primary features of CityGML are:

- Modularization
- Inter-operability of geo-spatial data

- Geometric representation of spatial data

- Semantics

- Topology

- Appearance

One of the main disadvantages of CityGML is that it's not as widely used for storing 3D models as other formats such as OBJ, 3DS, DAE and OSG are more popular. CityGML requires considerably large storage space compared to the more popular models that do not store semantic attributes. However, when compressed CityGML files like most XML file result in a very high compression ratio.

# LEVEL OF DETAILS DEFINED BY CITYGML 3D MODELS

CityGML incorporates multi-scale representation of 3D models with 5 Level of details(LOD). The LOD concept in CityGML is different from the one in computer graphics since it denotes the model's spatio-semantic adherence to its real-world counterpart [2].

- LOD 0 – It is essentially a 2.5d digital terrain model which are some mapped with an aerial image of the terrain. The semantics is modeled by a Building instance with corresponding attribute values [3].

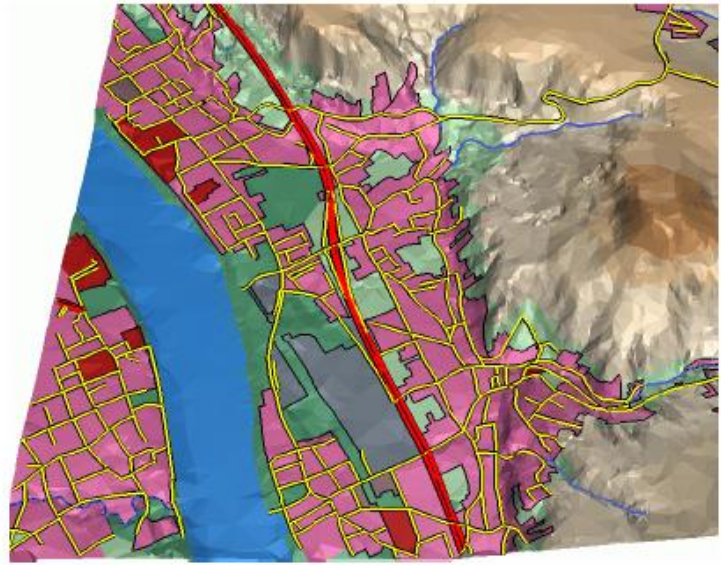

**Figure 1- LOD 0**

- LOD 1 – In LOD 1 block models are used to represent buildings without any roof structure. It is represented in blocks either as a solid or a multi-surface[3].

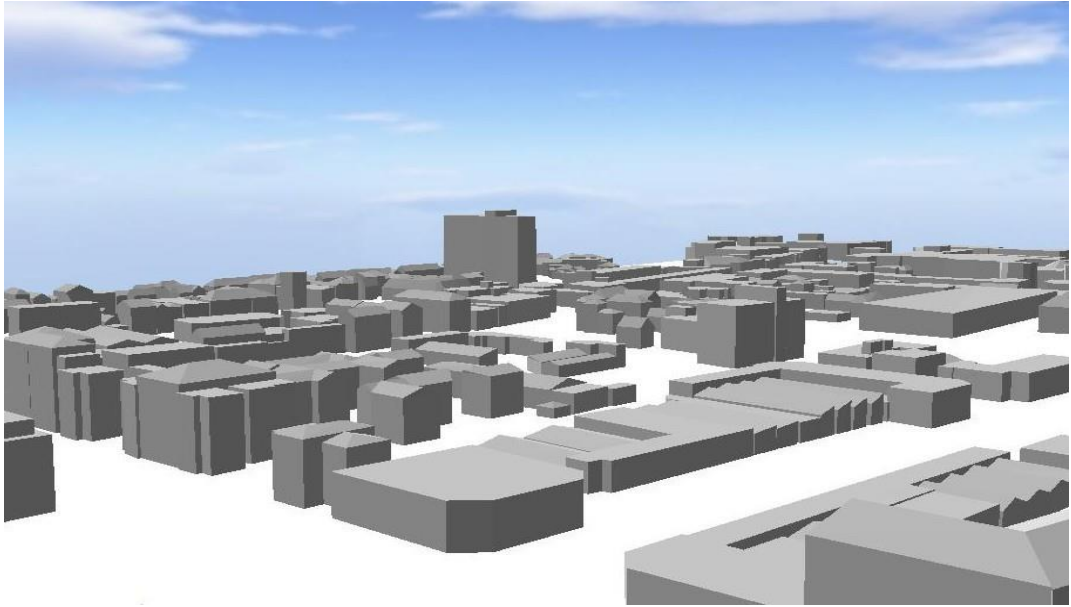

**Figure 2 - LOD 1**

- LOD 2 – In LOD 2 generalized roof structures are added, some thematic features such as wall surfaces and larger building installations such as balconies are also added. As openings are not supported in this LOD, closure surfaces are used to seal any openings present in the building structure. In version 2.0, two types of thematic surfaces, the OuterCeilingSurface and the OuterFloorSurface have been added to the list of thematic surfaces of the outer boundary of buildings[3].

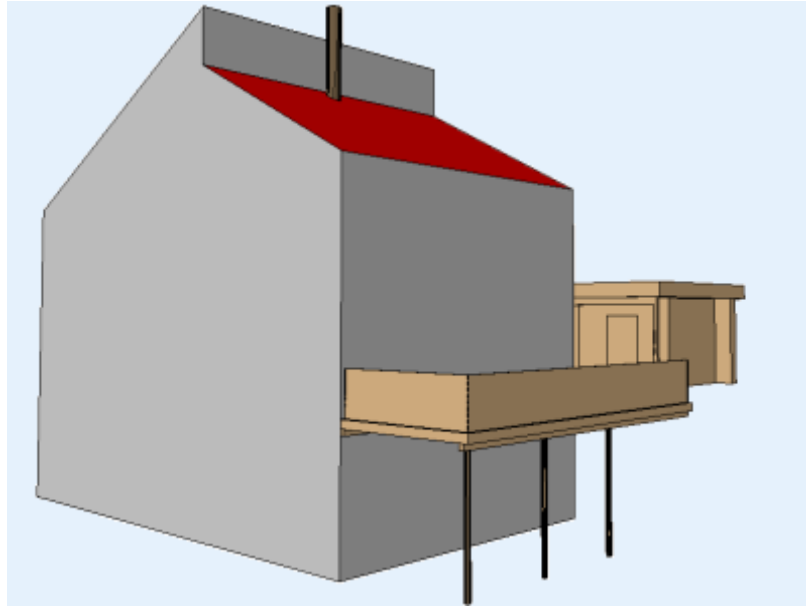

**Figure 3 - LOD 2**

- LOD 3 – In LOD 3 extends the model by adding openings such as doors, windows with detailed roof structures including building installations such as chimneys. Spatial properties of its feature are represented by their own surface geometry in LOD 3.

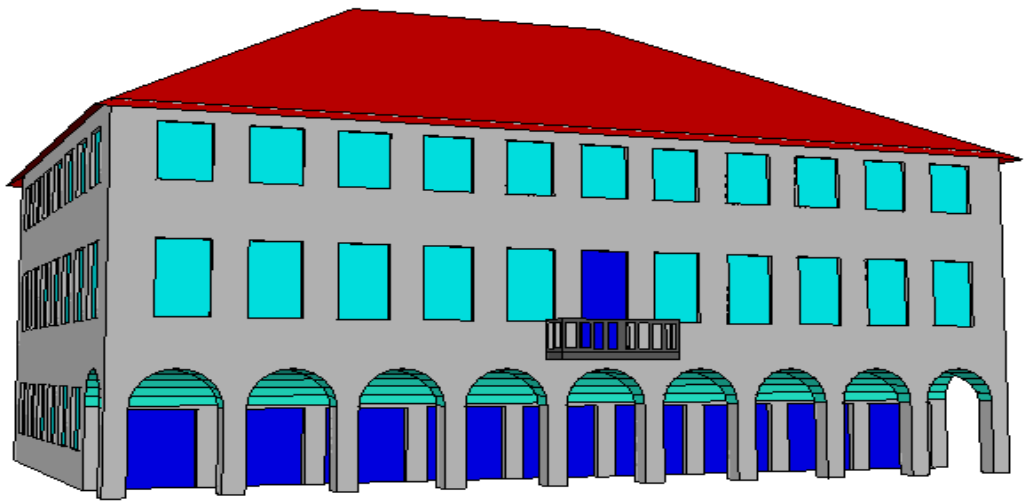

**Figure 4 - LOD 3**

- LOD 4 – In LoD4, in addition interior structures of buildings, tunnels and bridges etc. are added to the models. Several new features such as interior walls, building furniture's, rooms etc. are represented with their own geometry.

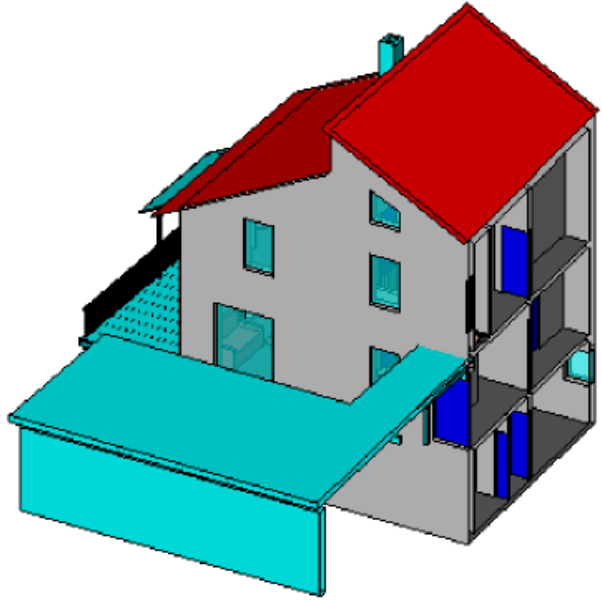

**Figure 5 - L0D 4**

# THEMATIC CLASSES OF CITYGML

CityGML covers all classes of object found in an urban area and they are not limited to buildings. CityGML contains several modules for each type of such objects including Building, CityFurniture, Transportation, Tunnel & Vegetation etc. The features are organized into modules (building module, vegetation module, transportation module, etc.), which can be arbitrarily combined as needed for a specific application[3].

A combination of CityGML modules is called a profile.

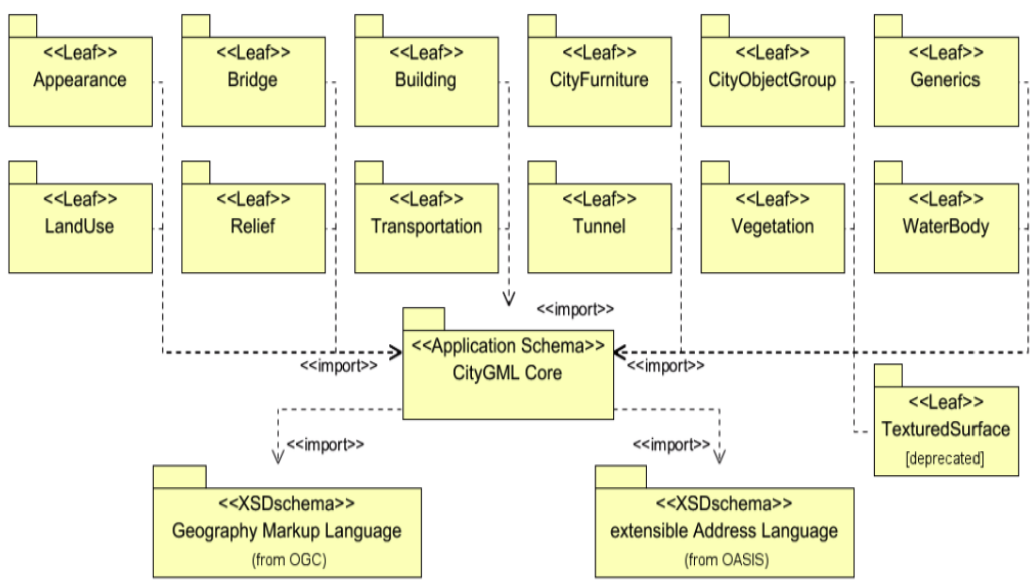

**Figure 6 - UML package diagram illustrating the separate modules of CityGML and their schema dependencies. Each extension module (indicated by the leaf packages) further imports the GML 3.1.1 schema definition in order to represent spatial properties of its them [4].**

- All the CityGML extension modules are based on the CityGML core module. The core module contains all the base classes; the thematic module classes are derived from these base classes. The basic data types which are not abstract are also defined in this module.

- The Appearance module is an Application Domain Extension of the abstract class _CityObject defined in the core module. They define the appearance features of the model such as material, textures which are the observable properties of the feature's surface.

- The Bridge module defines the representation of bridges, bridge installations, parts of bridges and the internal structure of bridges with both their thematic and spatial aspects in four levels of details from LOD 1 to LOD 4.

- The Building module defines the representation of buildings, building installations, parts of installations and the internal structure of installations with both their thematic and spatial aspects in five levels of details from LOD 0 to LOD 4. The building module is described in more details in the next sub-section.

- The CityFurniture module defines the representation of the city furniture objects which include benches, traffic signs, bus stops, lamp posts and advertisement banners among others.

- The CityObjectGroup module defines the aggregation of several different kinds of city objects into user-defined groups that can be used for the representation and transfer of a city area or a part of it. The group can be assigned special attributes that area defined by this module.

- The Generics module defines a generic extension to the CityGML data model that can be used to extend thematic classes, attributes and features that are not predefined by any other CityGML module. CityGML's Application Domain Extension mechanism is used augment the base class _CityObject to represent the additional generic attributes.

- The LandUse module is used to represent any earth surface dedicated to a specific type of land use.

- The Relief module defines the representation of the terrain in a city model in different levels of detail. The terrain may be specified as a regular raster or grid, as a TIN, by break lines, and by mass points [4].

- The Transportation module defines the transportation objects with a city model that includes roads, cars, buses and railways etc.

- The Tunnel module defines the representation of tunnels, tunnel installations, parts of tunnel and the internal structure of tunnels with both their thematic and spatial aspects in four levels of details from LOD 1 to LOD 4.

- The Vegetation module contains the thematic classes for representation of vegetation objects. The Vegetation module distinguishes between solitary vegetation objects like trees, & areas of vegetation like forests etc.

- The WaterBody module defines the representation of thematic and spatial aspects of rives, canals, lakes and basins. The do not represent the dynamic nature of these objects in the current specifications.

# BUILDING MODEL

The building more is the core class of CityGML object that is considered during this project to for conversion into CityGML. In this project the building models output as orientable, triangulated multi-surface surface meshes. The Building module defines the representation of buildings, building installations, parts of installations and the internal structure of installations with both their thematic and spatial aspects in five levels of details from LOD 0 to LOD 4.

The base class of the building model is the _AbstractBuilding class. _AbstractBuilding is sub-classed either by the Building class or the BuildingPart class. Building complexes with consists of a number of Buildings are aggregated using the CityObjectGroup class.

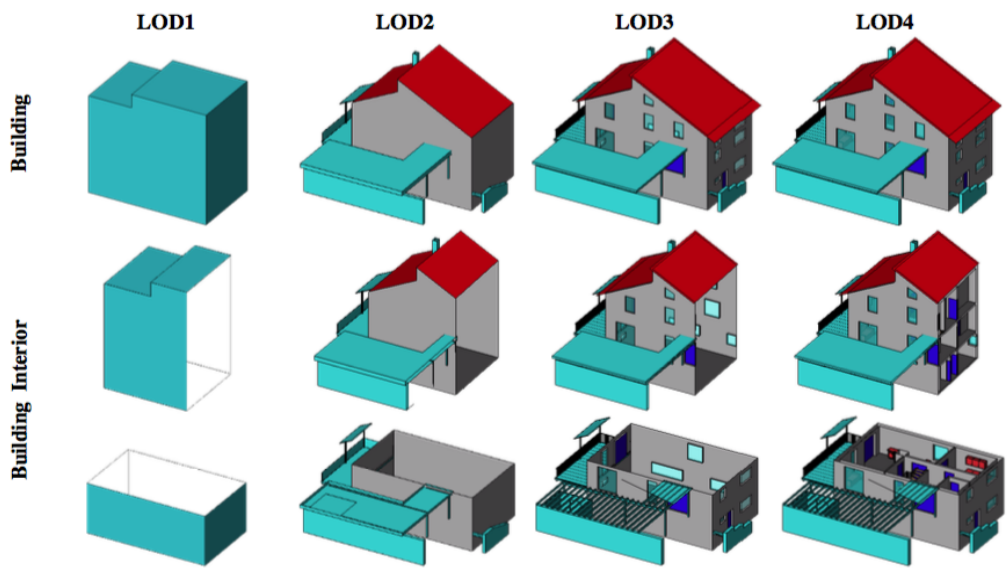

**Figure 7 - Building model in LOD1 – LOD4[4]**

The Building model is refined with additional features and attributes successively for each LOD ranging from LOD 0 to LOD 4. A building object can be represented simultaneously in different LOD's by providing separate geometric data for each LOD.

In LOD 0 horizontal surfaces and describing the building footprint and roof are used to represent a building object. The Surface geometries consists of vertices with 3D coordinates where all the vertices of a distinct surface have the same height values.

LOD 1 models are generalized geometric representations of the outer shells of building models and the various parts of the building are conjoined into a block and these parts are not represented separately. The geometric data can either be represented as lod1Solid or lod1MultiSurface.

In LOD 2, several new details are added to the external shell of the building model. These can be represented either as gml:Solid geometry or gml:MultiSurface geometry or a combination of the two. In this project the geometry is output as a gml:MultiSurface mesh. Features such as WallSurface, RoofSurface, OuterFloorSurface, OuterCeilingSurface and GroundSurface are also introduced in the LOD. ClosureSurface are used to seal off any openings in the geometry. BuildingInstallations are used to define parts of the building such as chimneys, antennas, balconies, external stairs and dormers.

LOD 3 models also define openings in the buildings such as doors and windows. According to GML3, the points have to be specified in reverse order (exterior boundaries counter-clockwise and interior boundaries clockwise when looking in opposite direction of the surface's normal vector) [4].

LOD 4 provides classes for representation of the interior building structure including Room, IntBuildingInstallation, FloorSurface, InteriorWallSurface, CeilingSurface, BuildingFurniture.

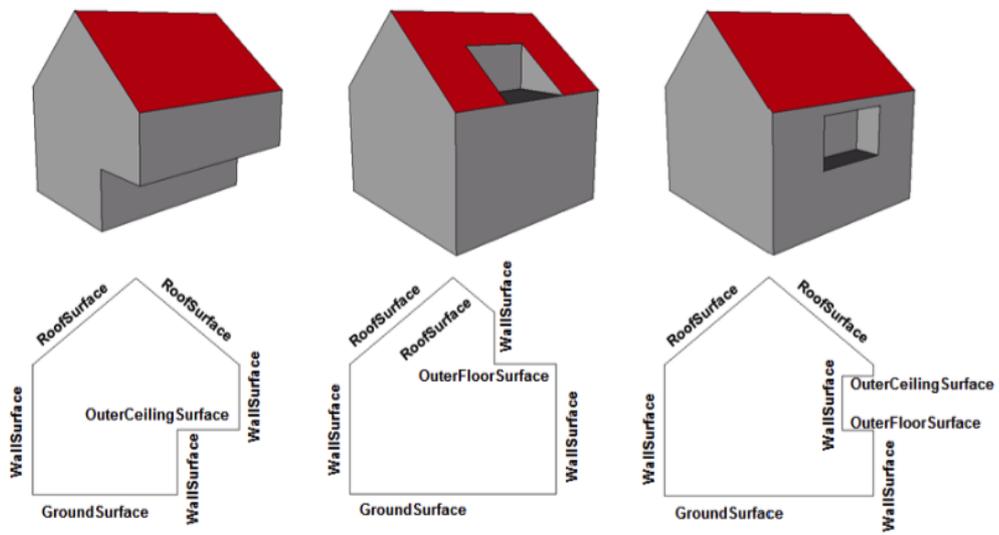

**Figure 8 - Examples of classification of external shell surfaces of a Building model *[4]*.**

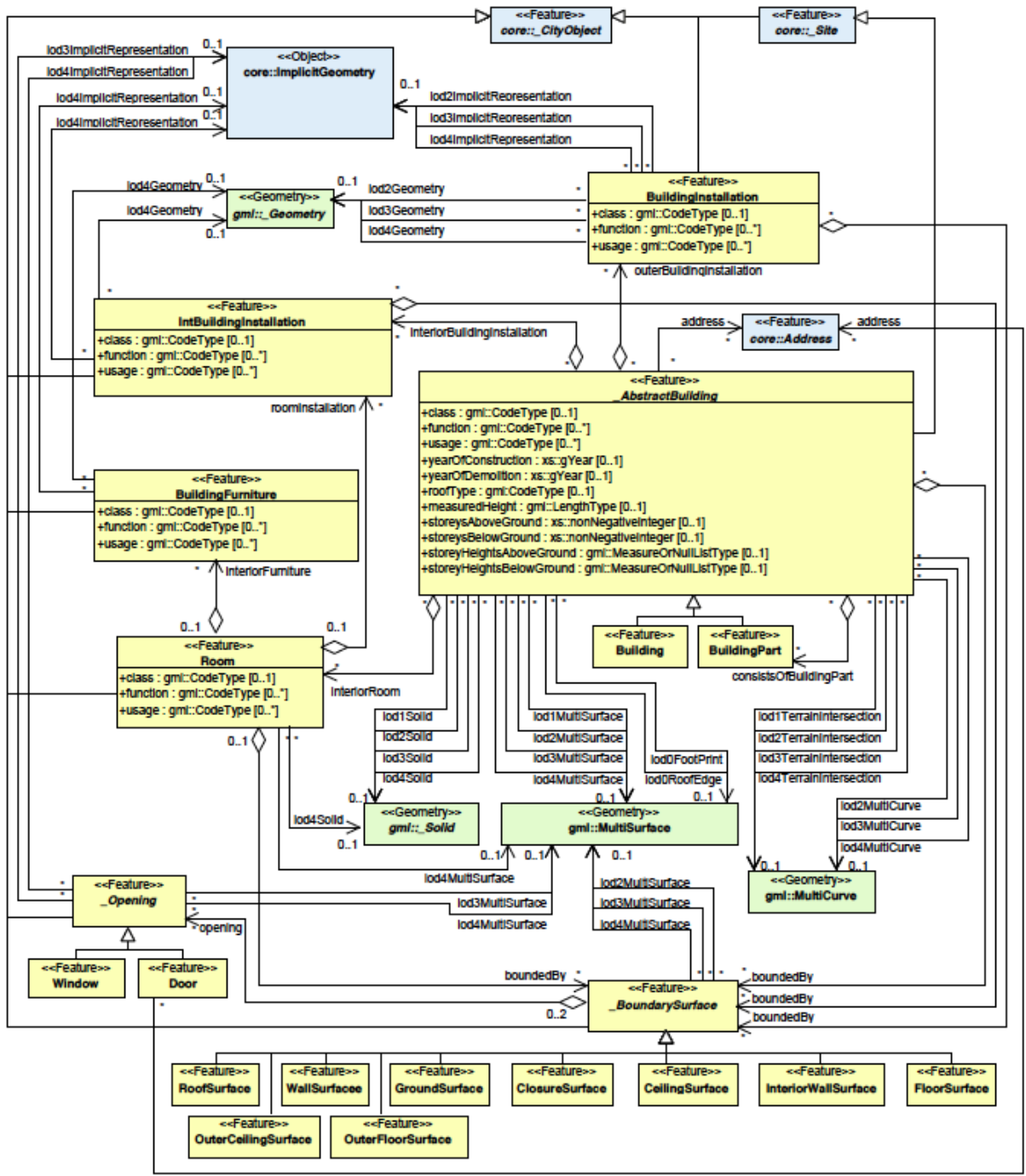

**Figure 9 - UML diagram of CityGML's building model. Prefixes are used to indicate XML namespaces associated with model elements. Element names without a prefix are defined within the CityGML Building module [4].**

# CITYGML VIEWERS

Several commercial and academic tools that visualize CityGML data are available both as commercial and free software. They support the visualization of sematic, geometric and spatial aspects of geo-visualization. The Aristoteles Viewer from the University of Bonn and the FZKViewer from KIT Karlsruhe are two such software.

1. The Aristoteles Viewer:

   The project is designed as a framework of applications to visualize GML 3D data. Aristoteles is an Open Source application developed by the the Institute for Cartography and Geoinformation, University of Bonn.

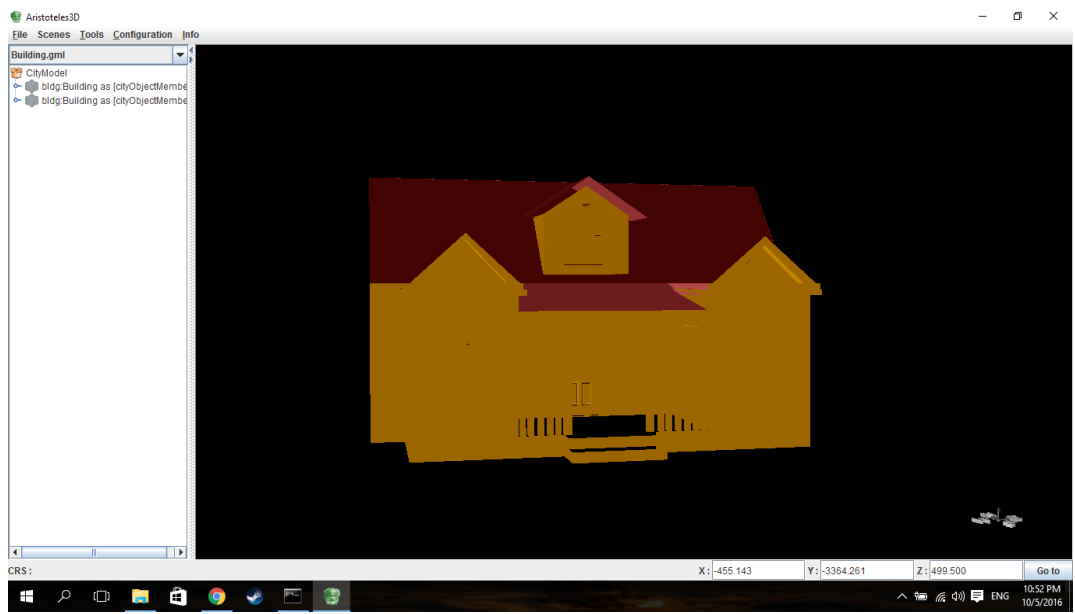

**Figure 10 - A screenshot of Aristoteles viewer**

2. The FZKViewer:

   This is a viewer used to visualize sematic data models like CityGML, IFC and gbXML developed by the Karlsruhe Institute of Technology.

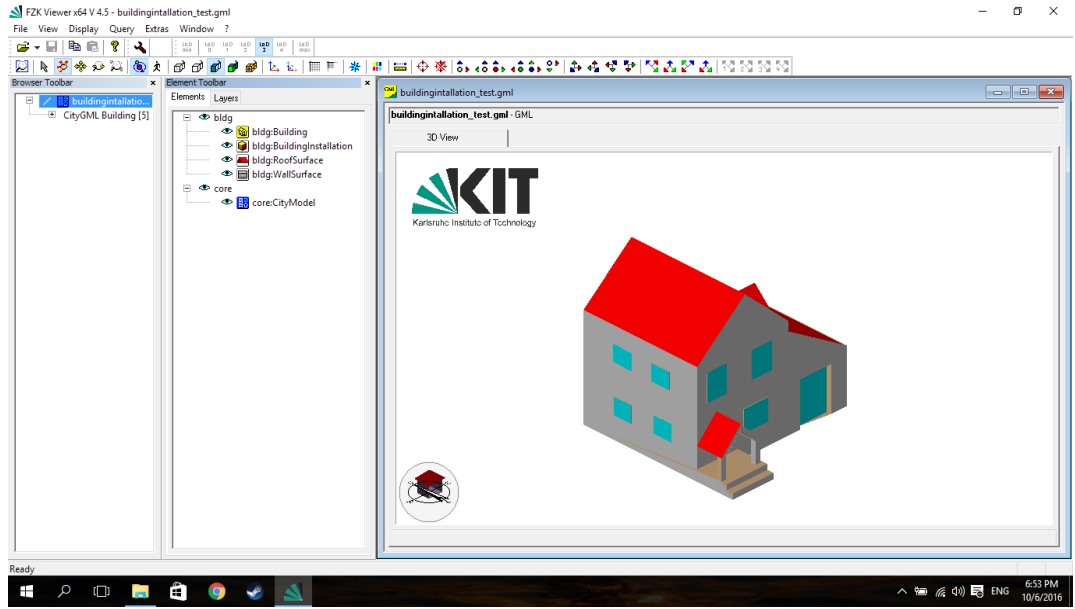

**Figure 11 - A screenshot of FZK Viewer displaying an output from OSG2CityGML prototype**

# OPENSCENEGRAPH

## WHAT IS A SCENE GRAPH?

Scene graph in general is a way of organizing data into a hierarchy of parent and child nodes. Any modification applied to the data in the parent node will also affect the data in the child node as the child node properties are relative to the properties of the parent node. Generally a tree data-structure is used for such representation. The scene is then generally drawn in a hierarchical manner starting at the root and moving downwards till the leaves. The nodes are arranged either representing spatial grouping, logical grouping or animation of objects. [5] have introduced OpenSceneGraph (OSG) as an open sourced, platform crossed high level graphics toolkit that frees the developer from implementing and optimizing low level graphics calls.

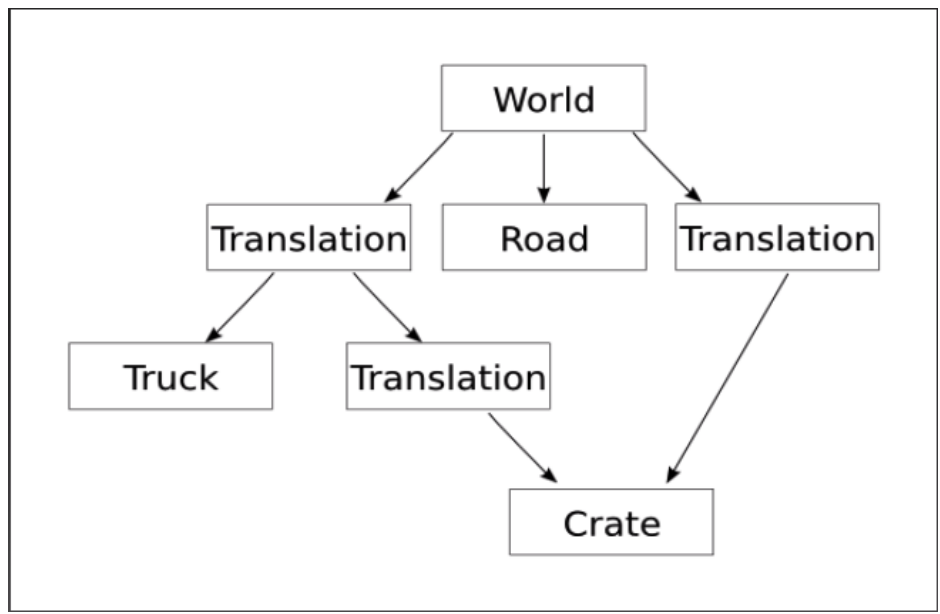

**Figure 12 - Scene graph for a scene with a truck on a road with one crate in the truck and another on the road [6].**

One example would be to create the car as a scene graph where the body is the parent and the wheels are child nodes. When updating the position of the parent node and the child nodes can be searched easily and their positions can also be updated.

Benefits of using a Scene Graph are:

1. Better performance: Arranging data in a managed hierarchical structure helps maximize the graphics performance.

2. Productivity: Helps increase the productivity as it reduces the time required to develop graphics applications.

3. Portability: Scene Graphs encapsulate lower level rendering thus when porting to another system the low level api changes need not be done by the developer.

4. Scalability: Scene Graphs can be easily scaled to draw a large scene with multiple objects with complex hierarchies which would be hard to manage without it.

# OVERVIEW OF OPENSCENEGRAPH

The OpenSceneGraph project was initiated by Don Burns in 1998.

**Components**

OpenSceneGraph provides libraries for producing highly scalable run-time functionalities. In addition to the core libraries a set of additional libraries have also been developed that handle specific aspects of application development. Below the OSG libraries used in this project are described briefly.

The core OSG libraries:

1. The osg library: This library is a collection of classes used to build scene graph. Some important classes are the Node class that is further sub-classed to classes such as Group and Geode that form the building blocks of the Scene Graph and scene management. The Drawable class is the base class that classifies geometric data and it extends the Node class. State classes are used to store the current state of rendering or to switch between rendering modes. Texture, material, light classes among others are also contained in this library to handle related functionalities. The library also contains math classes such that implement vector and matrix operations that are necessary to manipulate data in the three dimensional space. Some classes are used to manage pointer references and other scene management tasks.

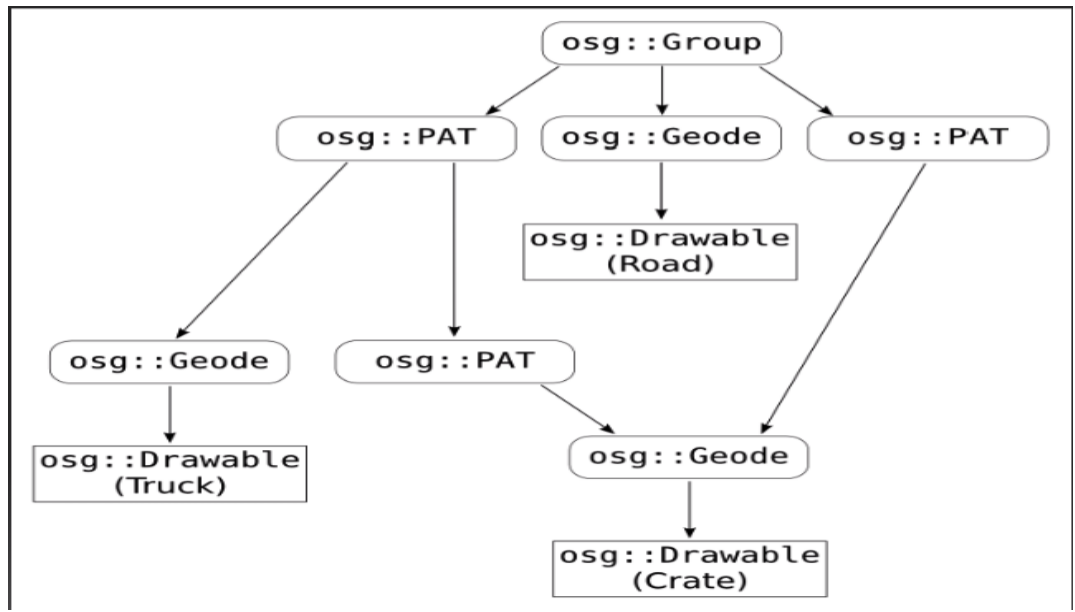

**Figure 13 - OSG representation of the same scene graph [6].**

2. The osgDB library: This library is quite important for the project as it provides a plugin mechanism to implement classes to load 3D mesh data from other file formats to the OpenSceneGraph format. The built in functionalities support read/write of files in OBJ, IVE, OSGT formats among others. About 45 plugins are available to read and/or write files to several formats that can be easily integrated into the project.

3. The osgUtil library: This library provides a set of utility classes and algorithms that can help traverse the nodes easily. Some classes implement the Visitor design pattern which is used to separate functions and algorithms from related data. These visitor classes then access the geometric data from the nodes and apply some modifications as defined in their function logic such as updating the color information. Some utility classes define the functionalities for ray-polygon intersections and polygon modifications etc.

4. The osgGA library: The osgGA library provides mechanisms for GUI interaction and abstraction.The classes handle interactive events from the peripheral devices such as keyboards, mouse etc.

5. The osgViewer library: It implements view-related classes that can be embedded into various windowing systems such as Qt, Windows, Cocoa etc.

Several other libraries are available with the OpenSceneGraph framework for specific functionalities that may not have been referenced in this project. These include osgTerrain, osgAnimation, osgFX, osgManipulator and osgText to name a few.

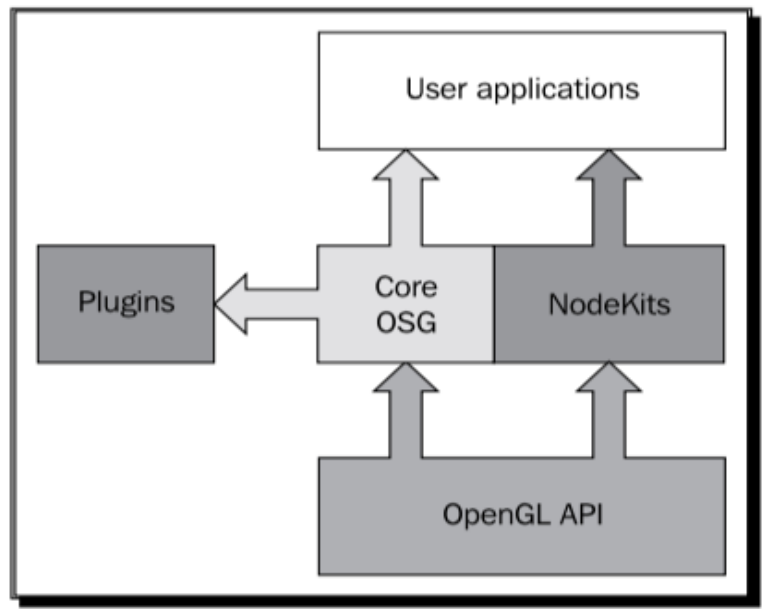

**Figure 14 - The OpenSceneGraph architecture *[7]*.**

# FORMAT CONVERSION

CityGML is an open data model and XML-based format for the representation and exchange of virtual 3D city models. It is based on the Geography Markup Language version 3.1.1 (GML3). Both CityGML and GML3 are international standards issued by the Open Geospatial Consortium (OGC) [8]. CityGML has been developed by the Special Interest Group 3D (SIG 3D) of the initiative Geodata Infrastructure North-Rhine Westphalia, Germany[8].

[2] investigated automatic semantic-preserving conversion between 3D models stored in OBJ format and to CityGML format. [9] developed a method for automatic repair of CityGML LOD2 buildings using shrink-wrapping. [3] give an overview of CityGML, its underlying concepts, its Levels-of-Detail, how to extend it, its applications, its likely future development, and the role it plays in scientific research.

## CityGML4J

CityGML4J is a java library that can be used to create CityGML objects [10]. This can be used to convert existing 3D model to CityGML file format by creating an GUI application that can utilize the CityGML4J library.

## IFC2CityGML

This is a software developed for automatic conversion of IFC objects to CityGML format IFC formats are used for architectural 3D modeling and they contain a lot of semantic information that can be mapped to the semantic definitions of CityGML format. [11] investigates the issues that occur in automatic conversion of 3D models in IFC format to CityGML format.

**SKETCHUP CITYGML PLUGIN**

This plugin for SketchUp allows generation of high quality 3D models in CityGML file format. The sematic information can be mapped with this plugin on to the model when it is being generated. It allows querying of geometry based on certain criteria from the Sketch file data and then the extracted geometry can be mapped to a CityGML class. However, it does not provide an inbuilt mechanism to convert from non SketchUp file formats such as OBJ to CityGML format. Selection of relevant surfaces based on simple queries are not very useful for models with high level of details.

# CHAPTER 3

# METHODOLOGY OF

# CONVERSION

## GEOMETRIC TRANSFORMATION

### TRIANGULATION

The main idea of converting various polygonal mesh types of the input geometric model to triangular meshes was uniform representation of geometries for easy segmentation and mapping of semantics. In this project the input meshes are first converted into triangle strip representation and then to a triangular mesh. This is further beneficial in generating linkages in the graph data structure between nearby vertices and faces which are not connected surfaces but lie close enough to be considered as a group.

Another advantages of converting the input mesh into a triangular mesh is that the CityGML output is uniform and thereby easy to validate. Triangulation simplified detection of shapes and co-planarity as fewer considerations were needed to develop such algorithms. A graph data structure is generated where each triangle face is wrapped by a Node. When two nodes share a vertex a link is generated between the two nodes. This graph is then used for traversal to compare the properties of each node.

# SPATIAL DATA-STRUCTURE TO EXTEND OPENSCENEGRAPH

## GRAPH GENERATION FOR PRE-PROCESSING AND SEMATIC MAPPING

The input mesh objects are normally multiple surfaces grouped as objects where two surfaces lie close to each other and appear to be one conjoined surface but if closely examines the they do not share any vertexes. In the modelling applications users are allowed to selected individual vertexes and meld them. However since CityGML supports such geometries asking the users to individually connect nearby vertices and weld them just for the purpose of segmentation and semantic mapping would be too much of hassle for the user. To simplify the task the geometries are wrapped with a graph data structure where each triangulated face is represented as a Node and the users have an option to create links between faces with vertices within a certain range from each other. The graph links thus generated are only for the purpose of graph traversal for identifying the shapes and semantics of the object. The underlying geometries are still not melded together and are intact.

A method for special partitioning of mesh data using buckets where the norm of the vector from origin to the vertex is the key and the vertex itself of the value is used to compare nearby vertices and link their parent faces.

The following steps are followed to generate a connected graph $G(V, E)$:

1. The users selects a level of precision $p$ in decimal places for the nearby vertices to be grouped together into the same bucket $B$.

2. The norm of the vector from origin $O$ to the Vertex $V$ is calculated, norm is rounded to the number of decimal places as set by the precision parameter. This rounded norm is then used as a key while the vertex is the value. All vertexes whose rounded distance to origin is the same as per the set precision end up in the same bucket. Therefore the bucket $B_n$ contains all the vertices whose norm $N$ when rounded off to the number of places as defined by precision $p$ equals $n$.

3. In the next step all the vertices in each bucket are compared, here the actual distances between the two vertices are calculated and if the distance is less than $\frac{1}{p}$, which is empirically derived then the links are created between the faces that they belong to (This is because two vectors with the same norm can still have a large angle between them).

Thus extra links in the graph are generated which were not generated in the original algorithm as the components of the graph $G$ was not connected. By default the algorithm for the connection of unconnected vertices is not run, only if the user chooses he can set the precision and generate an enhanced graph for analysis so he may using the existing connectivity in the mesh for segmentation of certain parts.

# SEMI-AUTOMATIC ASSIGNMENT OF SEMANTICS

## SEGMENTATION BASED ON FACE NORMAL

Normal based classification of semantics is mentioned by [2] where they deduce the semantic class of a surface from the orientation of the normal. Since the approach in this project is for semi-automated detection of semantics, the users are provided with a tool to easily segment triangular faces with normalized face normal within a close range.

The user is allowed to set a weight to approximate the similarity of the directions of the face normal(s) which may have slight variation due the process by which the mesh was generated.

Let $M$ be the mesh with a set of face $F_i$, $i = \{0 \ldots n\}$. Let be $s$ the index of the face selected by the user. Let $w$ be weight factor and $N_i$ be the normal of the face $F_i$.

$$SegNormal(M, N_s) = \sum_{i=0}^{n} F_i \quad \begin{cases} \|N_s - N_i\| <= w \\ -1.00 \geq w \geq 1.00 \end{cases}$$

**Figure 15 - Equation – 1**

$SegNormal(M, N_s)$ is the segmentation result of the mesh $M$ given a face normal $N_s$.

# SPATIAL SEGMENTATION BY GRAPH COMPONENT ANALYSIS

Segmentation of a component $C$ in the graph $G(V, E)$ is carried out. This is beneficial for two reasons, the first reason is that it allows the user to check if the entire region on the mesh is connected and the second reason is that if for some specific reasons like texture mapping etc. geometries are separated into multiple groups and aggregated to form the original mesh e.g. using a CAD software then the user can select a particular region of interest by selecting a node in the disconnected component of the graph that represents the entire mesh. Furthermore shape analysis algorithms can be built on top of the connected component analysis. The graph traversal starts from the selected node and every node that has a connectivity to the selected node is added to the segmentation.

*Let $M$ be the graph of the mesh with a set of faces which form the nodes of the graph. Let be $s$ the index of the face $F_s$ selected by the user.*

*Algorithm:*

1. *Input: $F_s$*

2. *Find the set of faces $F_i$, $i = \{0 \dots n\}$ that are first degree connections to the face $F_s$ and add all these faces to the list of selected faces.*

3. *Take each face $F_i$ in the set of connected faces from Step 2 as input and Goto: Step 1.*

4. *Stop when no more new faces can be found.*

# SEGMENTATION BASED ON SHAPE RECOGNITION

Some common shapes of surface meshes, if detected and segmented can aid the user greatly in easy semantic mapping of the mesh object. Thus using the graph data structure of connected faces, algorithms can be run for segmentation of a surface based on shape analysis:

A. Plane surface: In this co-planarity of triangular faces are analysed by checking the co-planarity of the centroids of the faces being tested with the vertices and the centroid of the original face selected by the user. The weight here helps select faces whose scalar triple product with the centroid of the selected plain is with a range of co-planarity (i.e. Scalar Triple Product = 0).

Let $M$ be the mesh with a set of face $F_i$, $i = \{0 \ldots n\}$. Let be $S$ the index of the face selected by the user. Let $W$ be weight factor and $C_i$ be the centroid of the face $F_i$. Let $U, V, W$ be the vertices of the face $F_s$ selected by the user. The * notation is for dot product and the ^ notation is for the cross product of vectors.

$$SegCoplanar(M, F_s) = \sum_{i=0}^{n} F_i \quad \begin{cases} (W - U) * ((V - U)^{Z-W}) \leq w \\ -1.00 \geq w \geq 1.00 \end{cases}$$

**Figure 16 - Equation – 2**

$SegCoplanar(M, F_s)$ is the segmentation result of the mesh $M$ given a face normal $N_s$.

B. Spatially connected plane surface: Segmentation of a component $C$ in the graph $G(V, E)$ is carried out along with the test for co-planarity. The algorithm stops when either all the nodes of the graph component have been visited or the new nodes on the paths being traversed are not coplanar to the original face $F_s$ selected by the user.

*Let $M$ be the graph of the mesh with a set of faces which form the nodes of the graph. Let be $S$ the index of the face $F_S$ selected by the user.*

<u>*Algorithm:*</u>

1. *Input: $F_s$*

2. *Find the set of faces $F_i$, $i = \{0 \ldots n\}$ that are first degree connections to the face $F_s$.*

3. *Take each face $F_i$ in the set of connected faces and test for co-planarity with $Equation-2$. Add all those faces that are coplanar to the Input face to the list of selected faces $\bar{F}$*

4. *Take each face $\bar{F}$ in the set of connected faces from Step 2 as input and Goto: Step 1.*

5. *Stop when no more new faces can be found.*

C. Wall surface: This segmentation aims to select the wall surfaces of a building while excluding the roof and other such surfaces.

A default UP vector which is the direction the normal of a plane roof surface is defined which the user may choose to change that. The UP vector that is used by the algorithm to exclude surfaces such as roof and floor surfaces. When the user selects a face, the linked faces in the graph whose normal are either perpendicular or parallel to the face normal are selected but whose faces whose normal are parallel the UP vector are not selected. The weight helps control if the UP vector is exactly parallel or within a certain range.

*Algorithm:*

1. *Input:* $F_s$

2. *Find the set of faces $F_i$, $i = \{0 \dots n\}$ that are first degree connections to the face $F_s$.*

3. *Take each face $F_i$ in the set of connected faces and test if their Normal are either perpendicular or parallel to the Normal of the selected face and still not parallel to the UP vector with $_{Equation - 3}$. Add all those faces that are coplanar to the Input face to the list of selected faces $\bar{F}$.*

Let $M$ be the mesh with a set of faces. Let be $s$ the index of the face selected by the user. Let $w$ be weight factor and $C_i$ be the centroid of the face $F_i$. Let $U, V, W$ be the vertices of the face $F_s$ selected by the user. The * notation is for dot product of vectors. Let $U$ be the UP vector of the mesh.

$$SegWall(M, N_s) = \sum_{i=0}^{n} F_i \quad \begin{cases} N_s * N_i \leq (0 + w) \cup N_s * N_i \geq (1 - w) \\ U * N_i \leq (1.0 - weight) \\ 0 \geq w \geq 1.00 \end{cases}$$

**Figure 17 - Equation – 3**

4. *Take each face $\bar{F}$ in the set of connected faces from Step 2 as input and Goto: Step 1.*

5. *Stop when no more new faces can be found.*

D. Curved surface: Curved surface segmentation is for selection of surfaces that are of curved or spherical topology. When the user selects a face then all the linked faces that have their face normal has a smooth gradient to the previous face in the iteration. The graph traversal stops once faces with whose gradient is not continuous as per the defined function are encountered.

Let $M$ be the mesh with a set of faces. Let be $s$ the index of the face selected by the user. Let $w$ be weight factor and $C_i$ be the centroid of the face $F_i$. Let $U, V, W$ be the vertices of the face $F_s$ selected by the user. The * notation is for dot product of vectors. Let $U$ be the UP vector of the mesh.

*Algorithm:*

1. *Input: $F_s$*

2. *Find the set of faces $F_i$, $i = \{0 \dots n\}$ that are first degree connections to the face $F_s$.*

3. *Take faces $F_i$ in the set of connected faces and test if their Normal are in a smooth gradient to the Normal of the input face $F_s$ with $_{Equation-4}$. Add all those faces that in a smooth gradient to the Input face to the list of selected faces $\bar{F}$.*

*Let $M$ be the mesh with a set of faces.. Let $S$ be the index of the face selected by the user.*

*Let $W$ be weight factor and $N_i$ be the normal of the face $F_i$. Let $N_s$ be the normal of the input face $F_s$. The * notation is for dot product of vectors.*

$$SegCurve(M, N_s) = \sum_{i=0}^{n} F_i \begin{cases} N_s * N_i < weight \\ 0 \geq w \geq 1.00 \end{cases}$$

**Figure 18 - Equation – 4**

4. *Take each face $\bar{F}$ in the set of connected faces from Step 2 as input and Goto: Step 1.*

5. *Stop when no more new faces can be found.*

E. Cylindrical surface: Cylindrical surface segmentation is for selection of surfaces that are of cylindrical topology. When the user selects a face then all the linked faces that have their face normal has a smooth gradient to the previous face in the iteration. The graph traversal stops once faces with whose gradient is not continuous as per the defined function are encountered. The defined function is different from the curved surface as it defines but a lower and upped bound for the gradient.

Let $M$ be the mesh with a set of faces. Let be $s$ the index of the face selected by the user. Let $w$ be weight factor and $C_i$ be the centroid of the face $F_i$. Let $U, V, W$ be the vertices of the face $F_s$ selected by the user. The * notation is for dot product of vectors. Let $U$ be the UP vector of the mesh.

*Algorithm:*

1. *Input: $F_s$*

2. *Find the set of faces $F_i$, $i = \{0 \ldots n\}$ that are first degree connections to the face $F_s$.*

3. *Take faces $F_i$ in the set of connected faces and test if their Normal are in a smooth gradient to the Normal of the input face $F_s$ with $_{Equation – 4}$. Add all those faces that in a smooth gradient to the Input face to the list of selected faces $\bar{F}$.*

Let $M$ be the mesh with a set of faces.. Let be $S$ the index of the face selected by the user.

Let $w$ be weight factor and $N_i$ be the normal of the face $F_i$. Let $N_s$ be the normal of the input face $F_s$. The * notation is for dot product of vectors.

$$SegCurve(M, N_s) = \sum_{i=0}^{n} F_i \begin{cases} N_s * N_i < weight \cap N_s * N_i < (1.0 - weight) \\ 0 \geq w \geq 1.00 \end{cases}$$

**Figure 19 - Equation – 5**

4. *Take each face $\bar{F}$ in the set of connected faces from Step 2 as input and Goto:*

   *Step 1.*

5. *Stop when no more new faces can be found.*

# CHAPTER 4

# PROTOTYPE, VALIDATION AND RESULTS

## OSG2CITYGML PROTOTYPE

The prototype is implemented in C++ using the Qt and OpenSceneGraph framework. The input model formats supported are .osg, .ive and .obj, while there is a possibility of adding other popular formats as the plugin implementation for these are available.

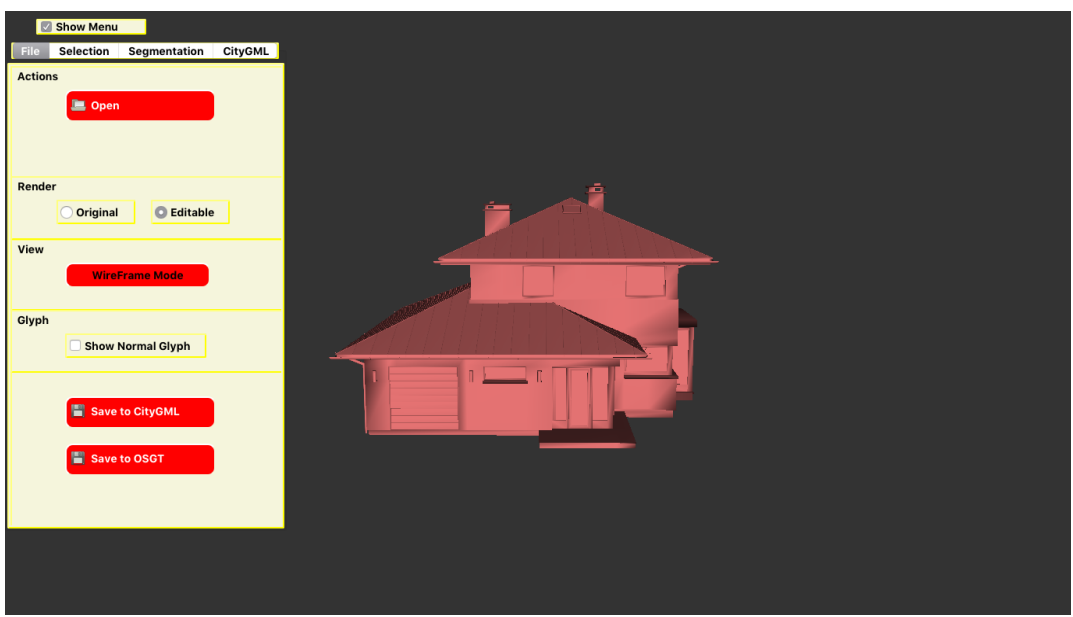

**Figure 20 - The prototype displaying a model in editable view**

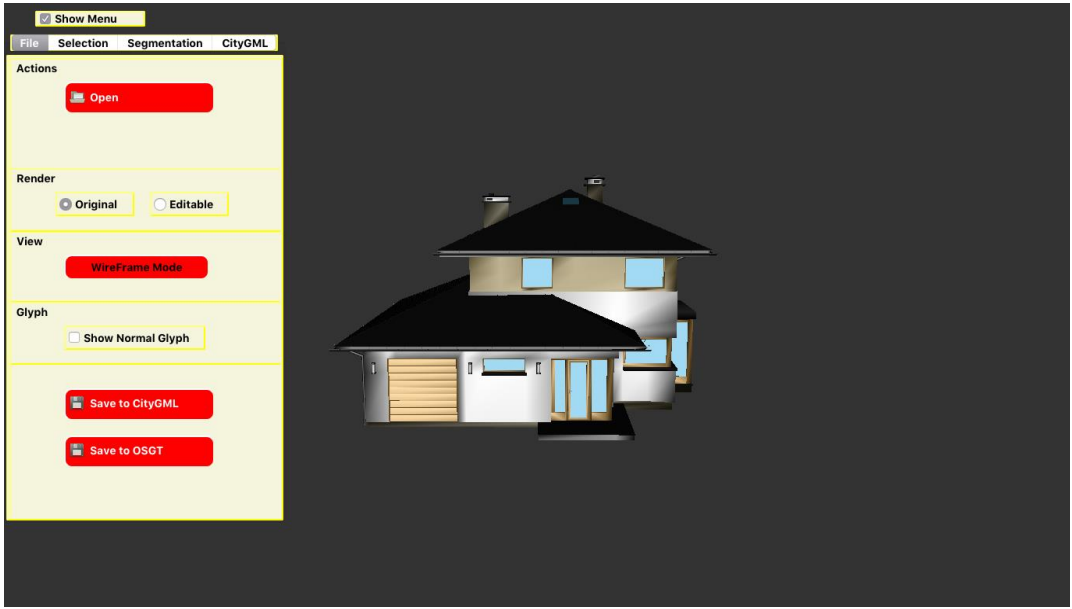

**Figure 21 - The prototype displaying a model in the shaded view**

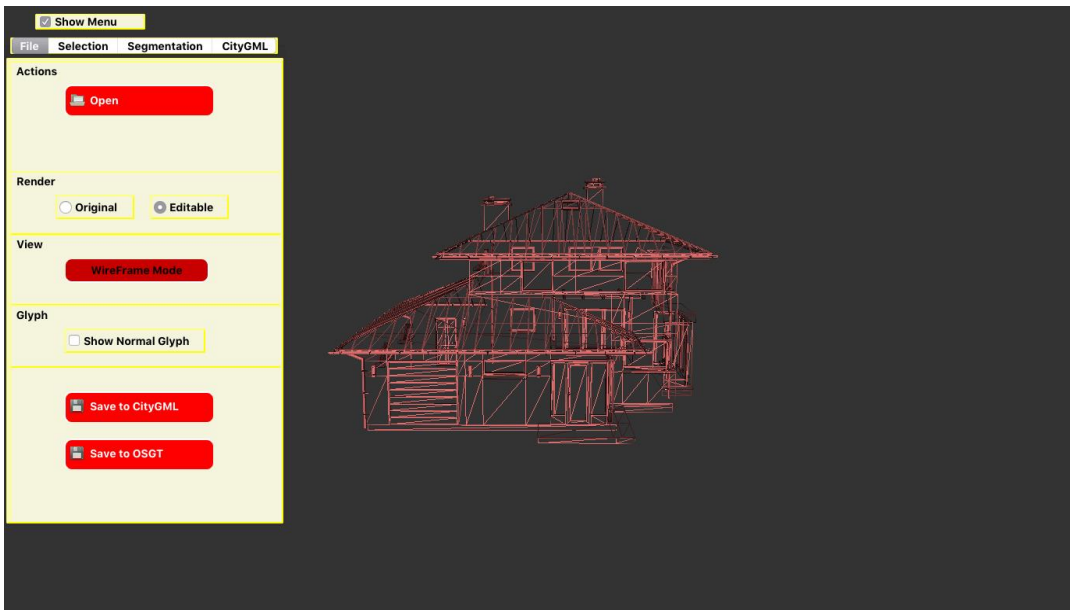

**Figure 22 - The prototype displaying the model in wireframe mode**

# TOOLS FOR EASY SEGMENTATION AND SEMANTIC MAPPING

## SEGMENTATION TOOLS

**Normal Segmentation**: This tool helps segment all surface in a model that have the same normal vector as mentioned in segment SEGMENTATION BASED ON FACE NORMAL.

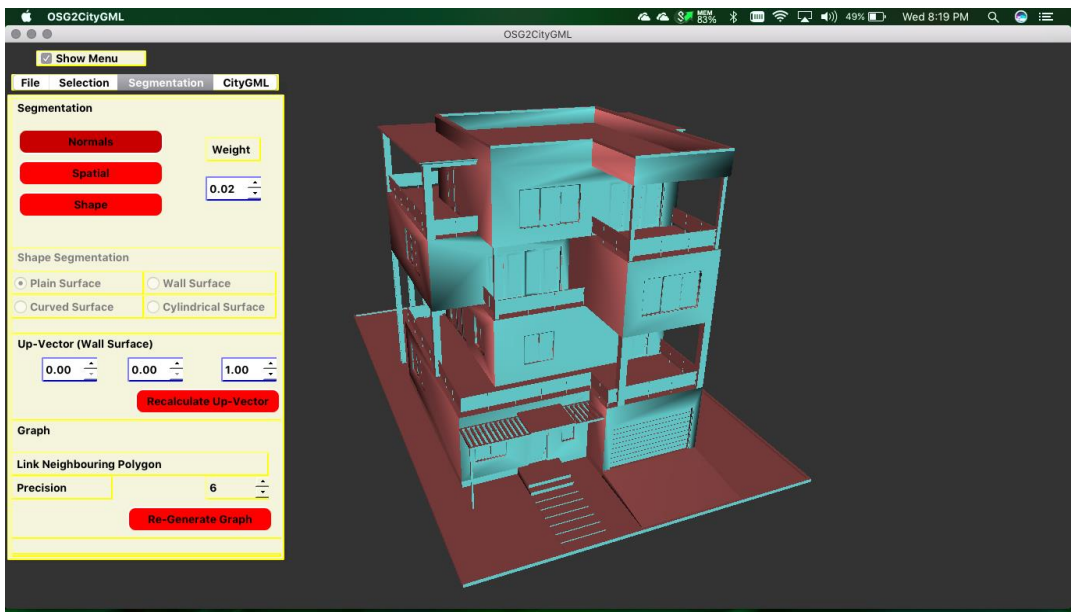

**Figure 23 - After applying the normal based segmentation all faces with normal pointing to the same direction as adjusted by the weight are segmented**

**Spatial Segmentation**: This tool helps segment a connected component of the mesh as mentioned in SPATIAL SEGMENTATION BY GRAPH COMPONENT ANALYSIS.

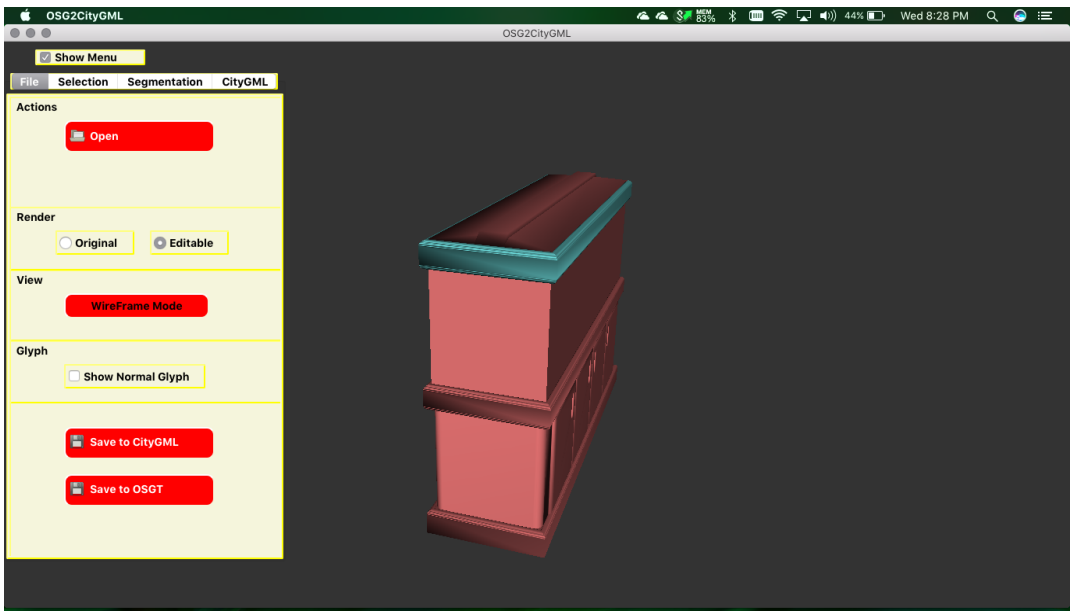

**Figure 24 - Spatial segmentation allows the selection of the connected component of a graph representing the underlying geometric mesh data**

Normal and Spatial segmentation tools can be used together to help segment connected faces of the mesh that have the same normal.

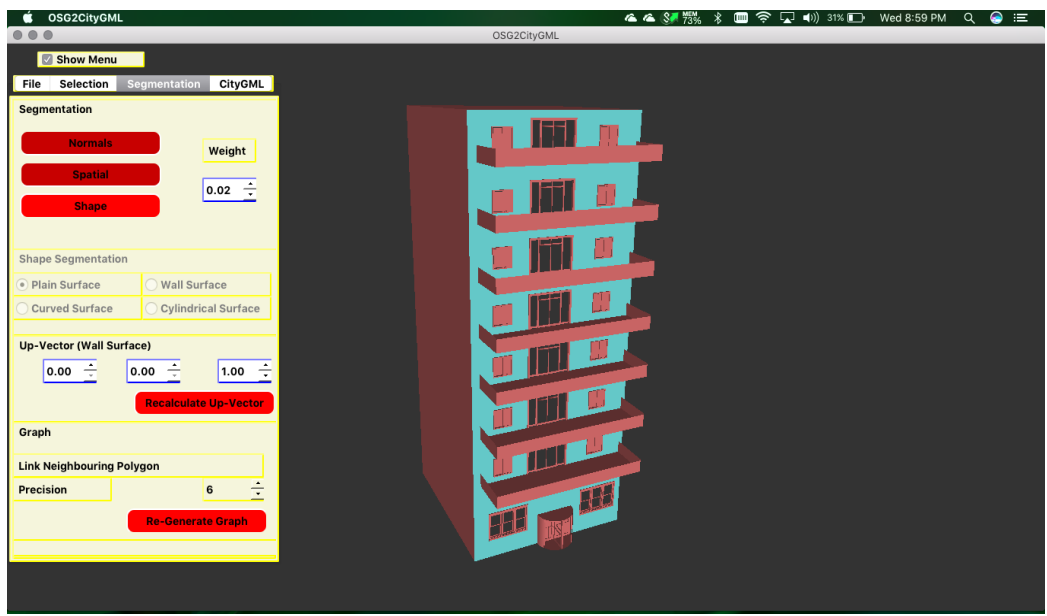

**Figure 25 - A combination of normal and spatial segmentation segments faces that are connected and whose normal point to the same direction as adjusted by the weight value**

**Shape Segmentation:** This is a set of tools that help segment surface in a model based on shape analysis.

- ***Plane surface:*** This tool helps segment a face of the mesh that are co-planar as mentioned in

- SEGMENTATION BASED ON SHAPE recognition section A

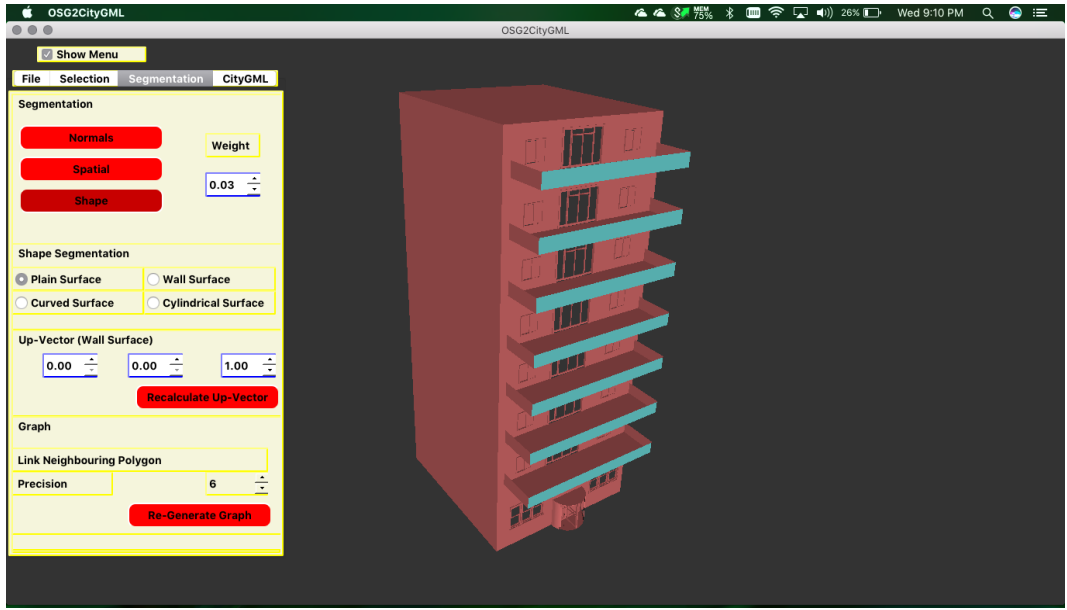

**Figure 26 - Plane based segmentation segments all the triangle primitives that are coplanar to the one selected by the user**

- ***Spatial & Plane surface:*** This tool helps segment a face of the mesh that are co-planar and spatially connected as mentioned in

- SEGMENTATION BASED ON SHAPE recognition section B

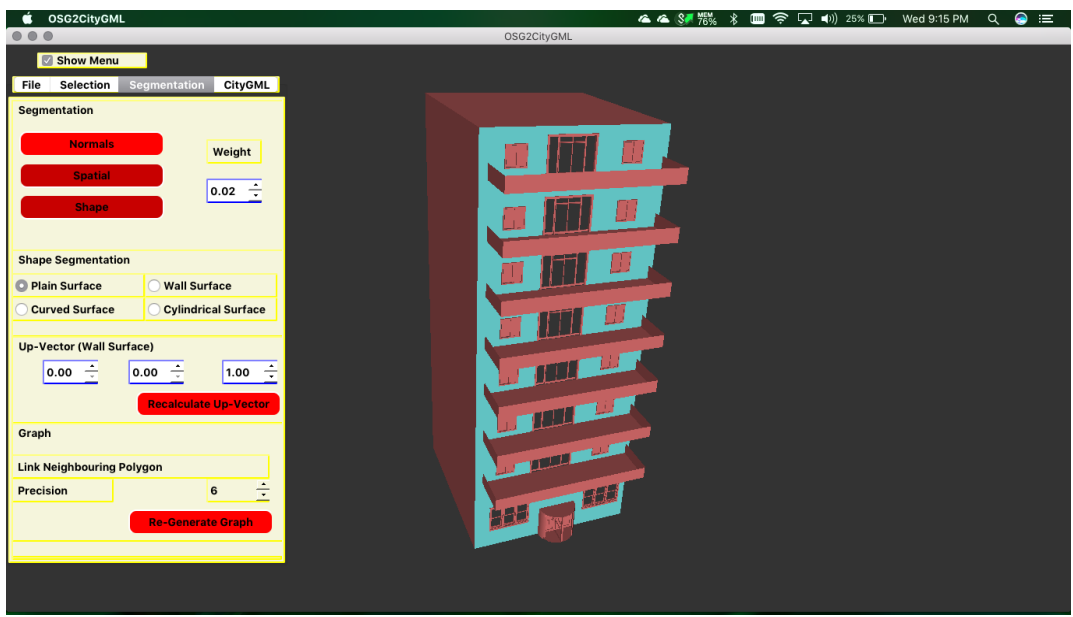

**Figure 27 - Spatial and Plane segmentation applied together**

- **Wall surface:** This tool helps segment a wall surface of the mesh that are spatially connected as mentioned in

- SEGMENTATION BASED ON SHAPE recognition section C.

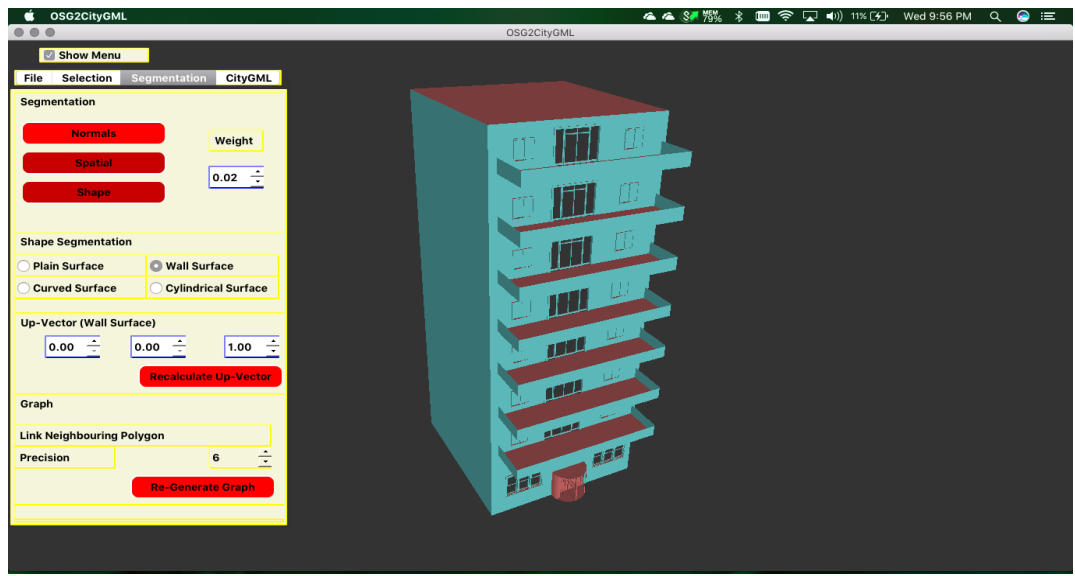

**Figure 28 - Wall segmentation (Wall surface segmentation tries to segment faces with orthogonal normal that are not parallel to the UP vector. This cannot differentiate between windows and walls surfaces or protruding balconies etc. if the normal are orthogonal to selected face.)**

- **Curved surface:** This tool helps segment a curved surface of the mesh that are spatially connected as mentioned in

- SEGMENTATION BASED ON SHAPE recognition section D

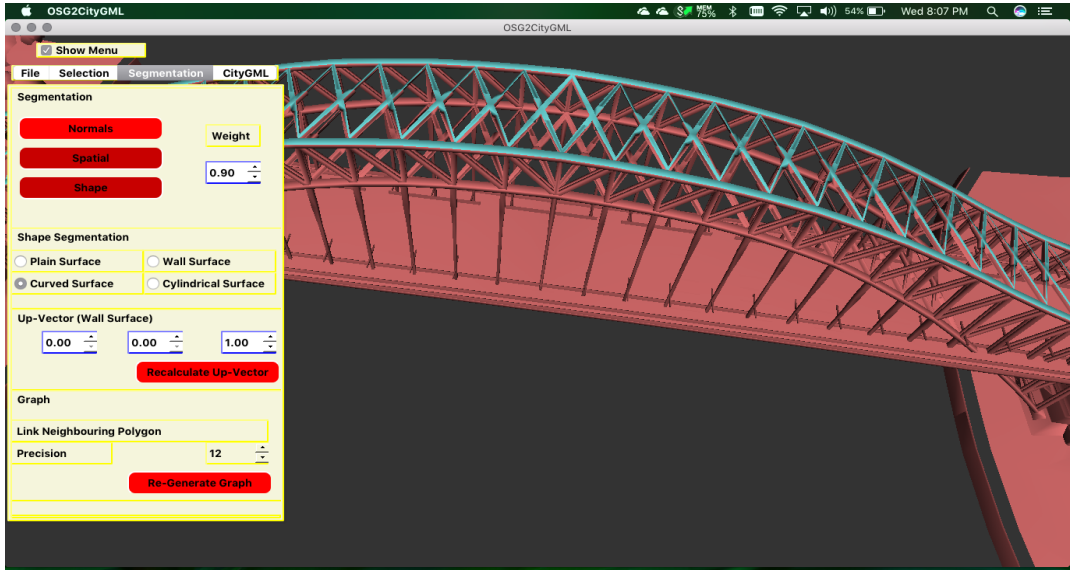

**Figure 29 - Curve segmentation (A bridge model has been used for illustration only although the scope of this project covers only building models. At weight equal to 0.90 only the top curved surface is selected)**

- *Cylindrical surface:* This tool helps segment a cylindrical surface of the mesh that are spatially connected as mentioned in

- SEGMENTATION BASED ON SHAPE recognition section E

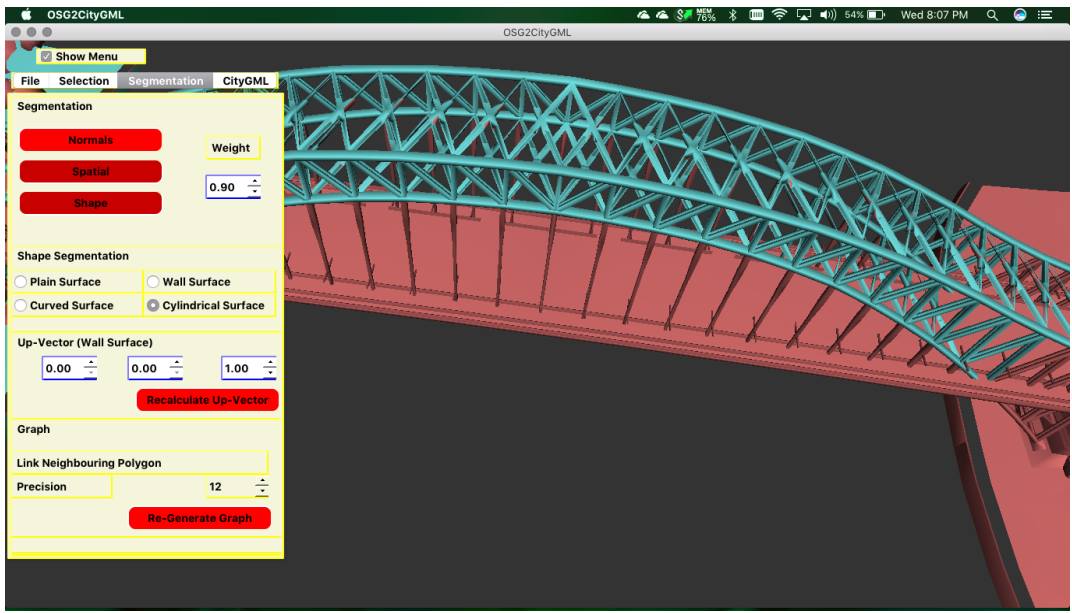

**Figure 30 - Cylindrical segmentation (A bridge model has been used for illustration only although the scope of this project covers only building models. At weight equal to 0.90 cylindrical surfaces are selected)**

# PAINT SELECTION TOOL

This tool has been provided for the user to paint selection/deselection over some regions that needs to be segmented manually. The user in Paint mode can click and drag over the surfaces, only the top surfaces in the ray – triangle intersection are selected.

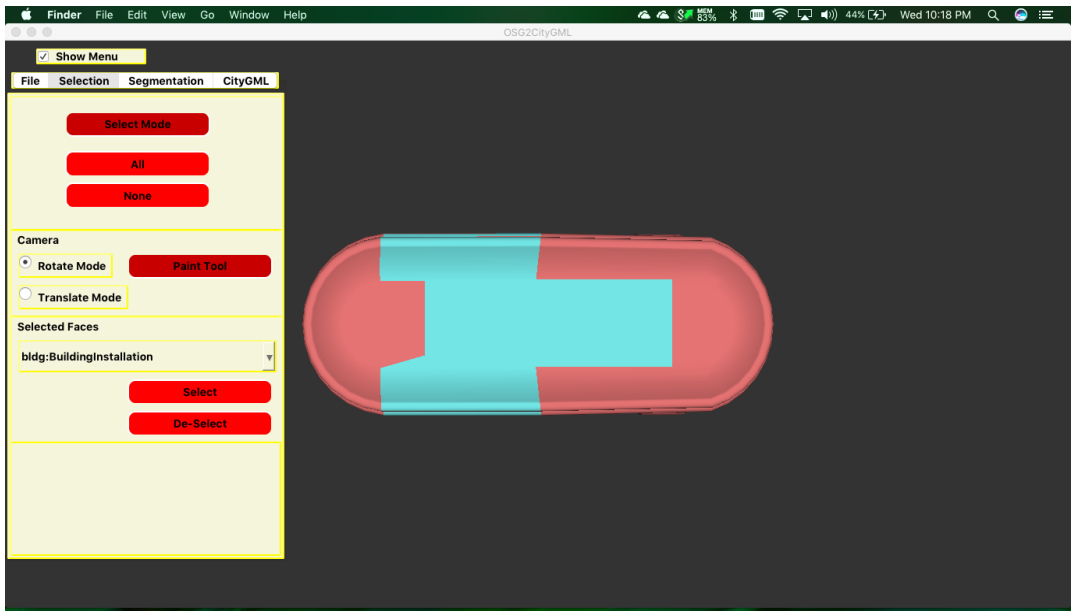

**Figure 31 - Paint selection tool**

# BINARY OPERATOR ON EXISTING SELECTION TOOL

This allows union or difference of previously selected and/or classified faces. Such combinations can be used to select the regions desired. Below is an example:

Step 1 - Select all

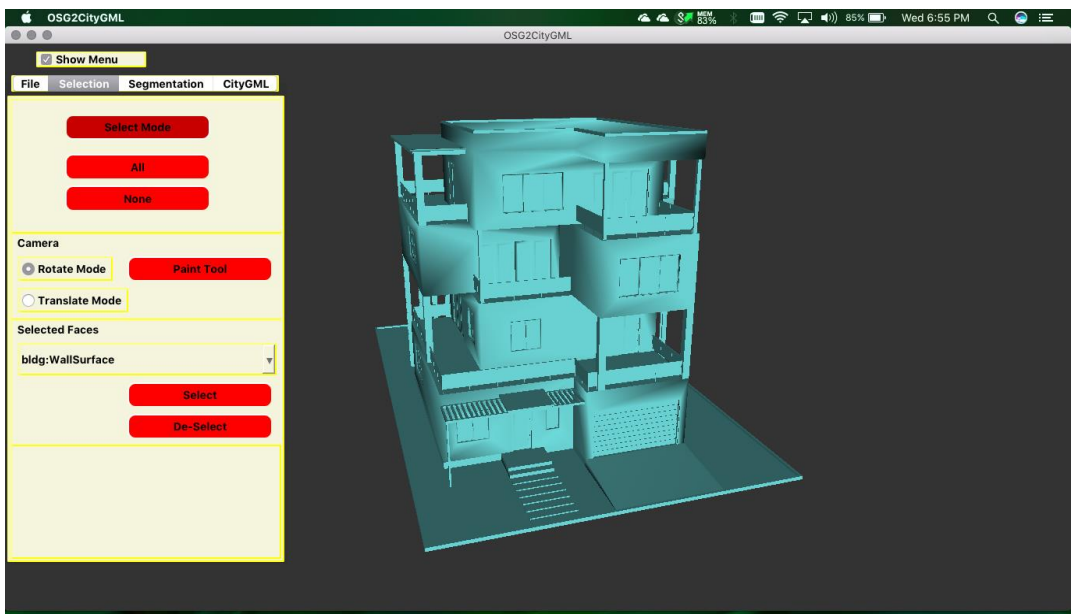

**Figure 32 - Entire mesh is selected**

Step 2 - Deselect the wall and roof

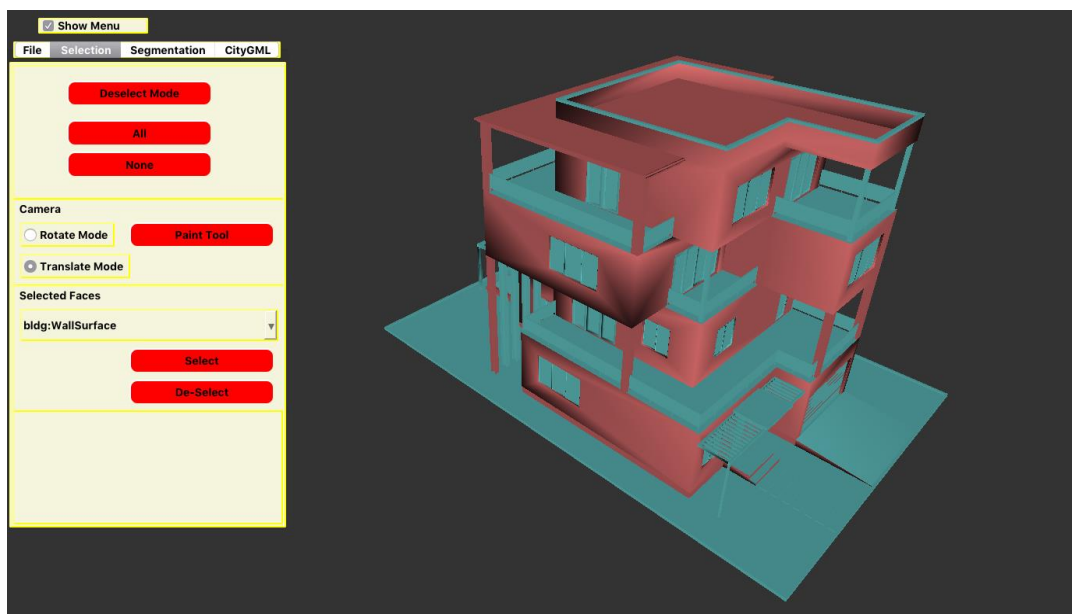

**Figure 33 - Deselect the wall and roof surfaces**

# CONNECTED GRAPH GENERATION TOOL

The tools helps regenerate a graph representation of the mesh while creating links between faces which do not share a vertex but their vertices are close enough as discussed in GRAPH GENERATION FOR PRE-PROCESSING AND SEMATIC MAPPING.

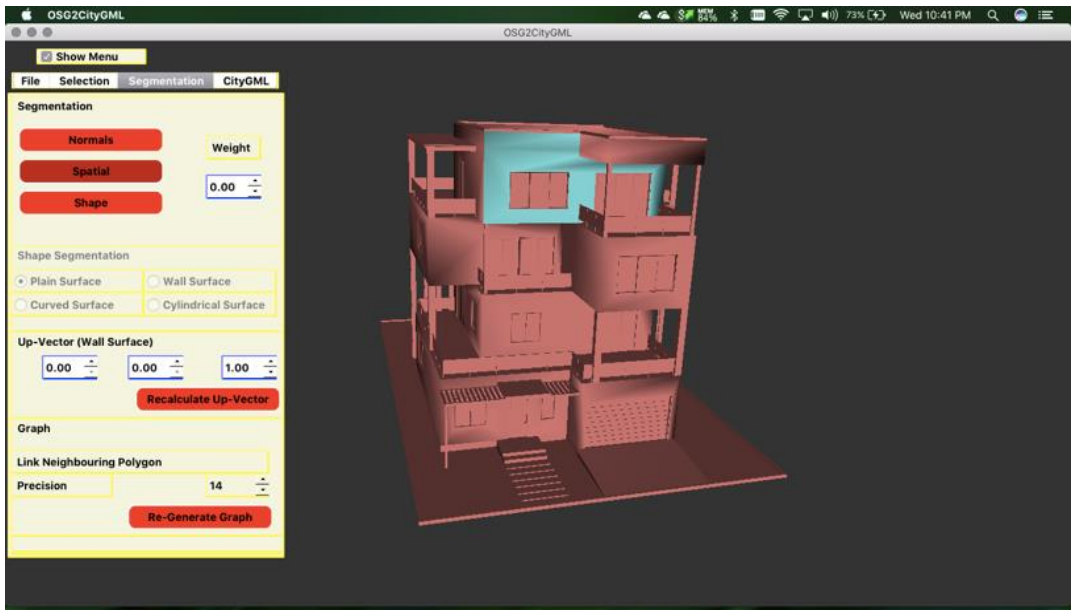

**Figure 34 - Figure 34 - Figure 34 - Spatial selection without graph regeneration**

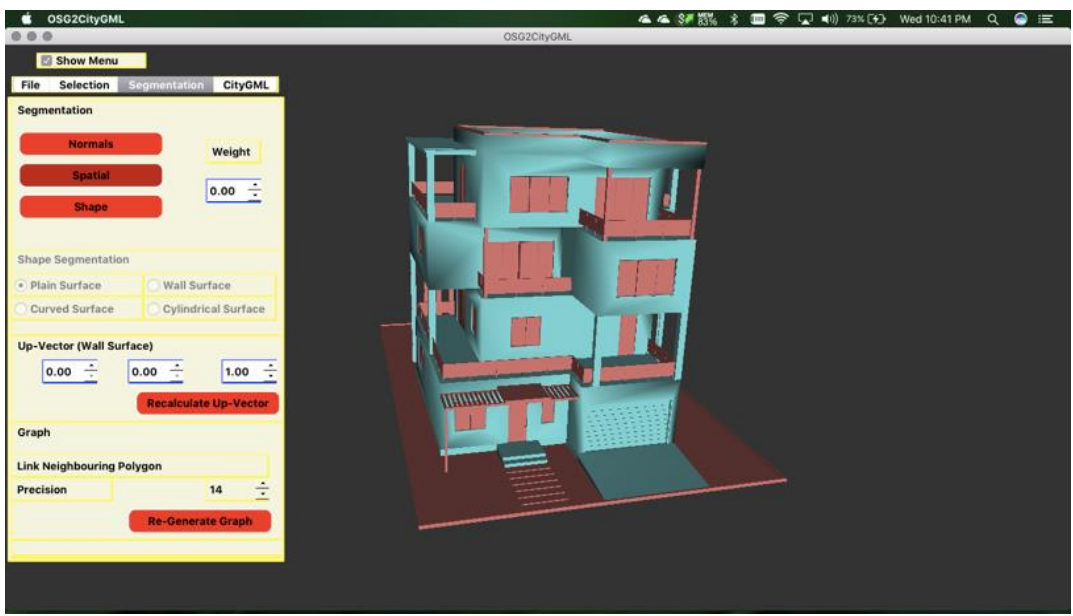

**Figure 35 - Spatial selection with graph regeneration at precision level 14, still some geometries are not connected**

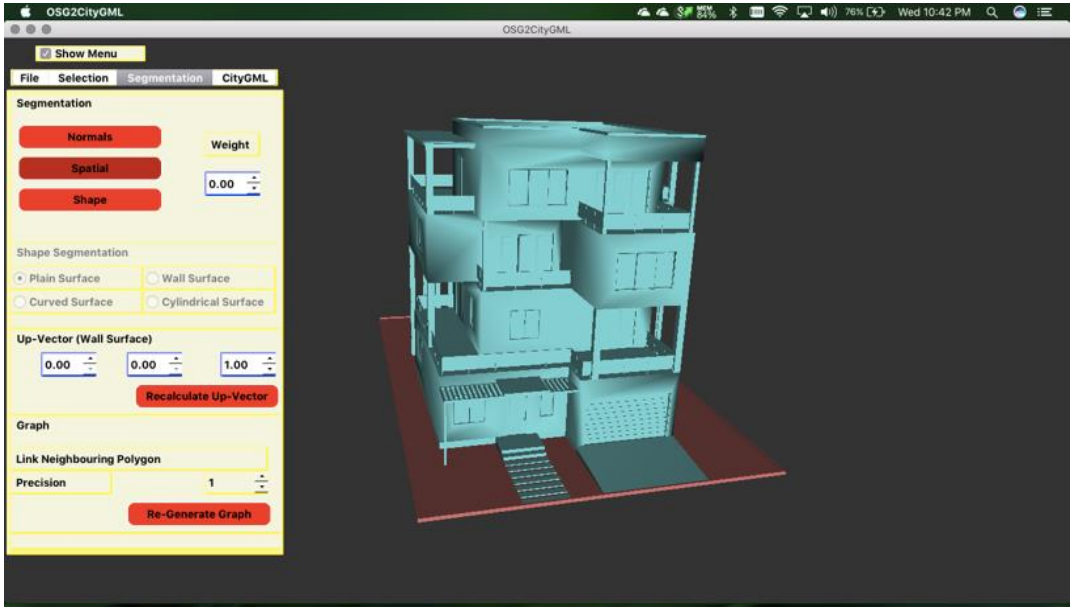

**Figure 36 - Spatial selection with graph regeneration at precision level 1**

# OUTPUT TO CITYGML FILE FORMAT

For the sake of consistency in output mesh in CityGML format a few project specific conventions were followed. The output is a multi-surface model with all primitives constrained to being triangle primitives. All the triangle primitives of a particular class are grouped together and output as a list of geometries of the particular type, this is just a convention followed in this project and this can be optimized to produced smaller files with faces grouped according to location rather than CityGML class.

A sample output:

```xml
<?xml version="1.0" encoding="UTF-8"?>
<CityModel                    xmlns="http://www.opengis.net/citygml/2.0"                           xmlns:gml="http://www.opengis.net/gml"
xmlns:xsi="http://www.w3.org/2001/XMLSchema-instance"                 xmlns:xAL="urn:oasis:names:tc:ciq:xsdschema:xAL:2.0"
xmlns:xlink="http://www.w3.org/1999/xlink"                    xmlns:smil20lang="http://www.w3.org/2001/SMIL20/Language"
xmlns:smil20="http://www.w3.org/2001/SMIL20/"                       xmlns:app="http://www.opengis.net/citygml/appearance/2.0"
xmlns:bldg="http://www.opengis.net/citygml/building/2.0"               xmlns:brdg="http://www.opengis.net/citygml/bridge/2.0"
xmlns:frn="http://www.opengis.net/citygml/cityfurniture/2.0"           xmlns:core="http://www.opengis.net/citygml/citygmlbase/2.0"
xmlns:grp="http://www.opengis.net/citygml/cityobjectgroup/2.0"           xmlns:gen="http://www.opengis.net/citygml/generics/2.0"
xmlns:luse="http://www.opengis.net/citygml/landuse/2.0"                  xmlns:dem="http://www.opengis.net/citygml/relief/2.0"
xmlns:tex="http://www.opengis.net/citygml/texturedsurface/2.0"        xmlns:tran="http://www.opengis.net/citygml/transportation/2.0"
xmlns:tun="http://www.opengis.net/citygml/tunnel/2.0"                  xmlns:veg="http://www.opengis.net/citygml/vegetation/2.0"
xmlns:wtr="http://www.opengis.net/citygml/waterbody/2.0"             xsi:schemaLocation="http://www.opengis.net/citygml/relief/1.0
http://schemas.opengis.net/citygml/relief/1.0/relief.xsd                              http://www.opengis.net/citygml/landuse/1.0
http://schemas.opengis.net/citygml/landuse/1.0/landUse.xsd                              http://www.opengis.net/citygml/building/1.0
http://schemas.opengis.net/citygml/building/1.0/building.xsd                    http://www.opengis.net/citygml/cityobjectgroup/1.0
http://schemas.opengis.net/citygml/cityobjectgroup/1.0/cityObjectGroup.xsd          http://www.opengis.net/citygml/cityfurniture/1.0
http://schemas.opengis.net/citygml/cityfurniture/1.0/cityFurniture.xsd                http://www.opengis.net/citygml/appearance/1.0
http://schemas.opengis.net/citygml/appearance/1.0/appearance.xsd                 http://www.opengis.net/citygml/texturedsurface/1.0
http://schemas.opengis.net/citygml/texturedsurface/1.0/texturedSurface.xsd          http://www.opengis.net/citygml/transportation/1.0
http://schemas.opengis.net/citygml/transportation/1.0/transportation.xsd               http://www.opengis.net/citygml/waterbody/1.0
http://schemas.opengis.net/citygml/waterbody/1.0/waterBody.xsd                        http://www.opengis.net/citygml/vegetation/1.0
http://schemas.opengis.net/citygml/vegetation/1.0/vegetation.xsd                         http://www.opengis.net/citygml/generics/1.0
http://schemas.opengis.net/citygml/generics/1.0/generics.xsd">
    <gml:description>OSG2CITYGML</gml:description>
    <gml:name>ModernHouse.gml</gml:name>
    <core:cityObjectMember>
      <bldg:Building>
        <bldg:outerBuildingInstallation>
          <bldg:BuildingInstallation>
            <bldg:lod3Geometry>
              <gml:MultiSurface>
                <gml:surfaceMember>
                  <gml:Polygon>
                    <gml:exterior>
                      <gml:LinearRing>
                        <gml:pos>-124.189 -258.724 12</gml:pos>
                        <gml:pos>-119.199 -258.724 16.99</gml:pos>
                        <gml:pos>-91.232 -258.724 12</gml:pos>
                        <gml:pos>-124.189 -258.724 12</gml:pos>
                      </gml:LinearRing>
                    </gml:exterior>
                  </gml:Polygon>
                </gml:surfaceMember>
              </gml:MultiSurface>
            </bldg:lod3Geometry>
          </bldg:BuildingInstallation>
        </bldg:outerBuildingInstallation>
        <bldg:outerBuildingInstallation>
          <bldg:BuildingInstallation>
            <bldg:lod3Geometry>
              <gml:MultiSurface>
```

```xml
                <gml:surfaceMember>
                  <gml:Polygon>
                    <gml:exterior>
                      <gml:LinearRing>
                        <gml:pos>-119.199 -258.724 16.99</gml:pos>
                        <gml:pos>-96.222 -258.724 16.99</gml:pos>
                        <gml:pos>-91.232 -258.724 12</gml:pos>
                        <gml:pos>-119.199 -258.724 16.99</gml:pos>
                      </gml:LinearRing>
                    </gml:exterior>
                  </gml:Polygon>
                </gml:surfaceMember>
              </gml:MultiSurface>
            </bldg:lod3Geometry>
          </bldg:BuildingInstallation>
        </bldg:outerBuildingInstallation>

• • •

        <bldg:opening>
          <bldg:Window>
            <bldg:lod3MultiSurface>
              <gml:MultiSurface>
                <gml:surfaceMember>
                  <gml:Polygon>
                    <gml:exterior>
                      <gml:LinearRing>
                        <gml:pos>-238.027 36.6589 102.13</gml:pos>
                        <gml:pos>-238.027 65.8565 102.13</gml:pos>
                        <gml:pos>-238.027 36.6589 46.1303</gml:pos>
                        <gml:pos>-238.027 36.6589 102.13</gml:pos>
                      </gml:LinearRing>
                    </gml:exterior>
                  </gml:Polygon>
                </gml:surfaceMember>
              </gml:MultiSurface>
            </bldg:lod3MultiSurface>
          </bldg:Window>
        </bldg:opening>
        <bldg:opening>
          <bldg:Window>
            <bldg:lod3MultiSurface>
              <gml:MultiSurface>
                <gml:surfaceMember>
                  <gml:Polygon>
                    <gml:exterior>
                      <gml:LinearRing>
                        <gml:pos>-238.027 65.8565 102.13</gml:pos>
                        <gml:pos>-238.027 65.8565 46.1303</gml:pos>
                        <gml:pos>-238.027 36.6589 46.1303</gml:pos>
                        <gml:pos>-238.027 65.8565 102.13</gml:pos>
                      </gml:LinearRing>
                    </gml:exterior>
                  </gml:Polygon>
                </gml:surfaceMember>
              </gml:MultiSurface>
            </bldg:lod3MultiSurface>
          </bldg:Window>
        </bldg:opening>
      </bldg:Building>
    </core:cityObjectMember>
</CityModel>
```

# RESULTS FROM THE PROTOTYPE

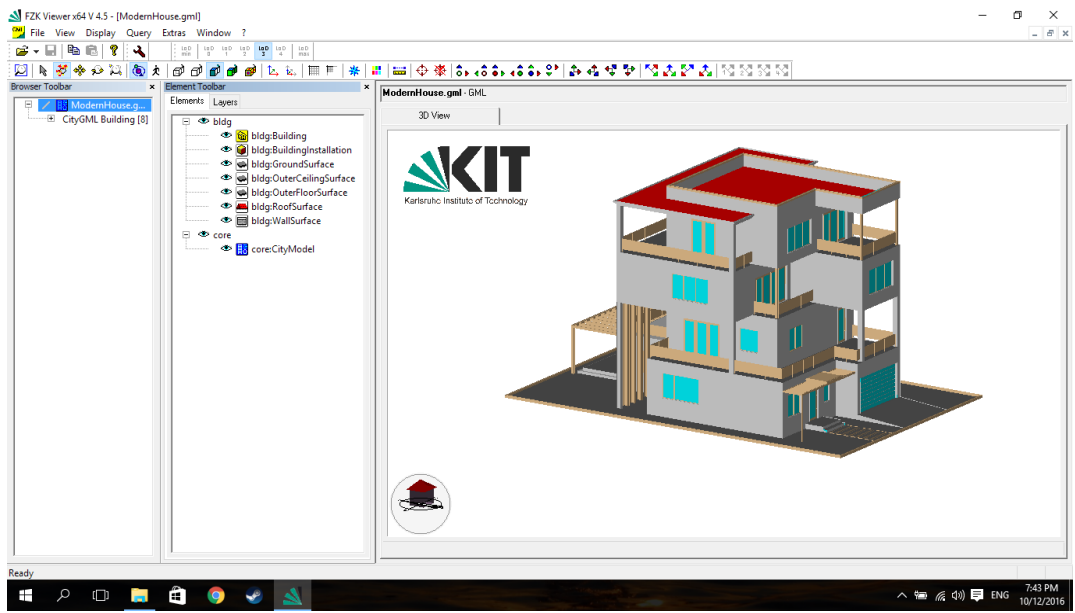

**Figure 37 - A Modern Building: This is an output from OSG2CITYGML prototype developed for this project.**

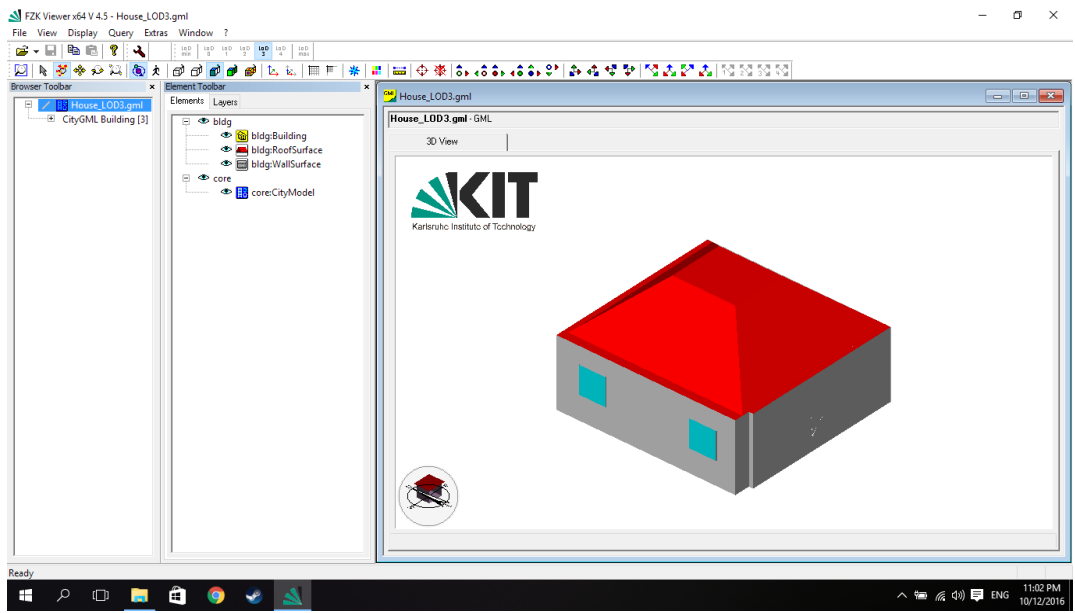

**Figure 38 - A small house: This is another output from OSG2CITYGML prototype developed for this project.**

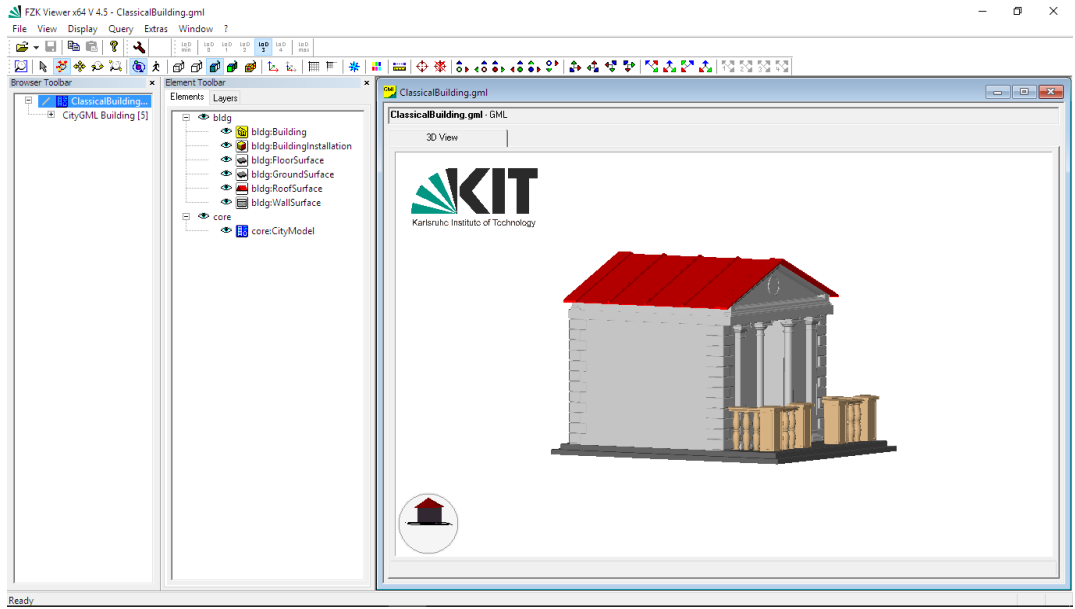

**Figure 39 - A Classical Building**

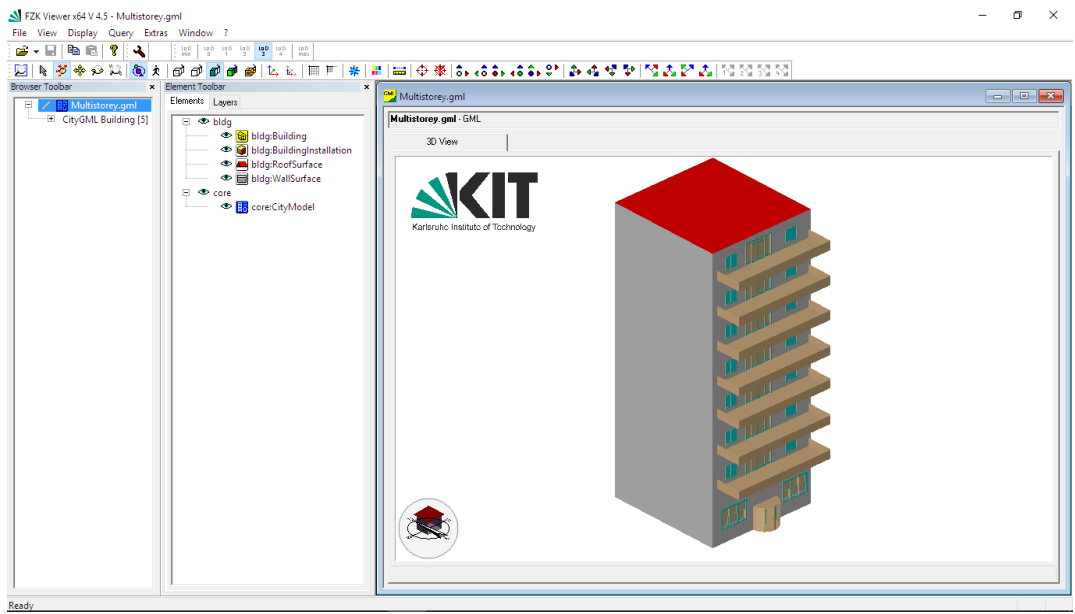

**Figure 40 - A Multi-storey Building**

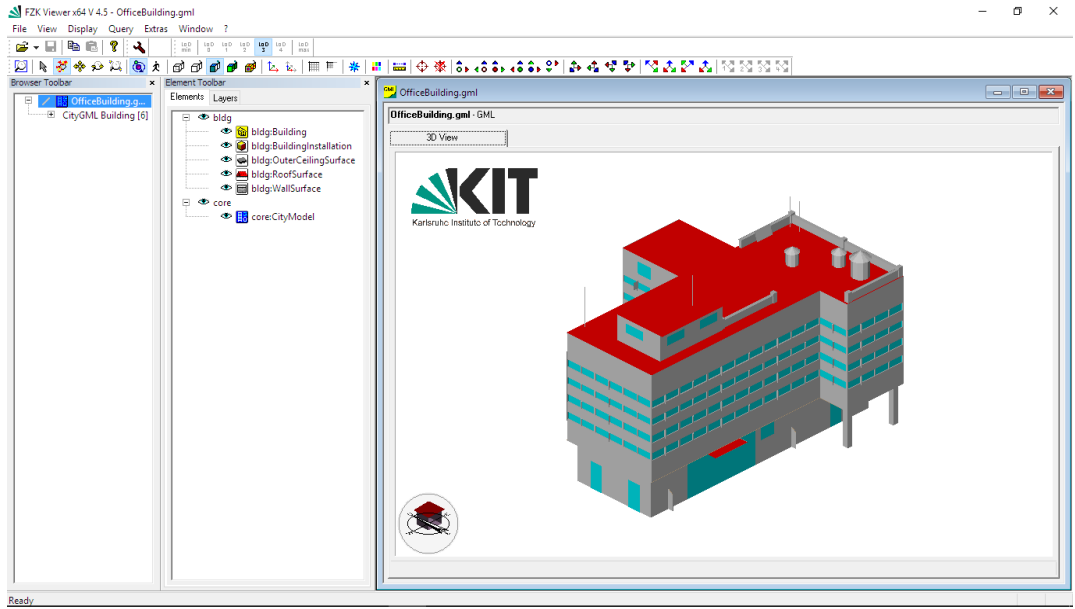

**Figure 41 - An office building**

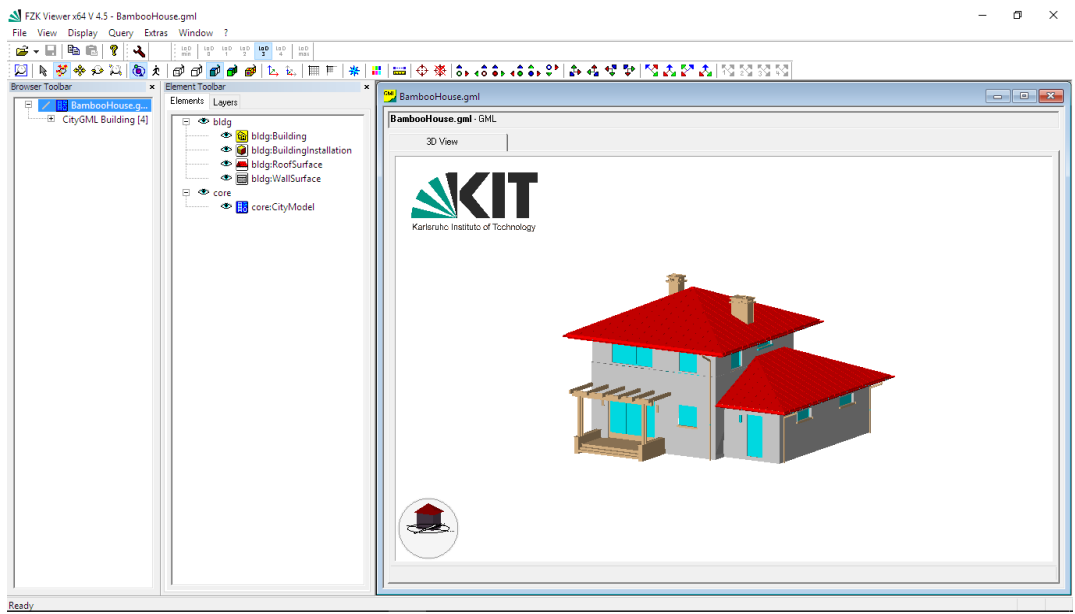

**Figure 42 - A Bamboo House**

# LIMITATIONS OF THE APPROACH

The output from the prototype have been rendered using FZK Viewer and validated using val3dity online validation tool. The error codes defined as per CityGML standards are the following:

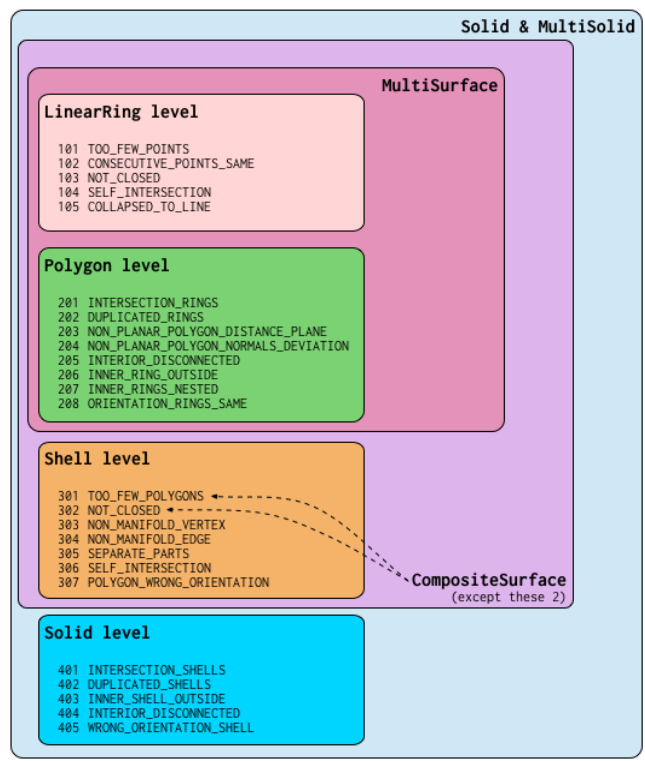

Figure 43 - Error codes for the validation of geometric primitives. [12]

While several measures have been taken to decrease the occurrence of validation errors, however a few common validation errors are encountered such as 104—SELF_INTERSECTION. However the prototype doesn't provide any method of highlighting or repair of such issues that already exist in the input model. In general these issues do not cause any major problems and the models are visualized properly using the CityGML viewers.

Automatic mapping of semantics has been discussed by [12] with the recommendations for classifications based on normal given in the diagram below. This automatic semantic mapping algorithm has not been included in the prototype as classes such building installation have varied normal and the classes such as floor & roof surfaces and outer ceiling & ground surfaces etc. can have normal pointing to the same directions.

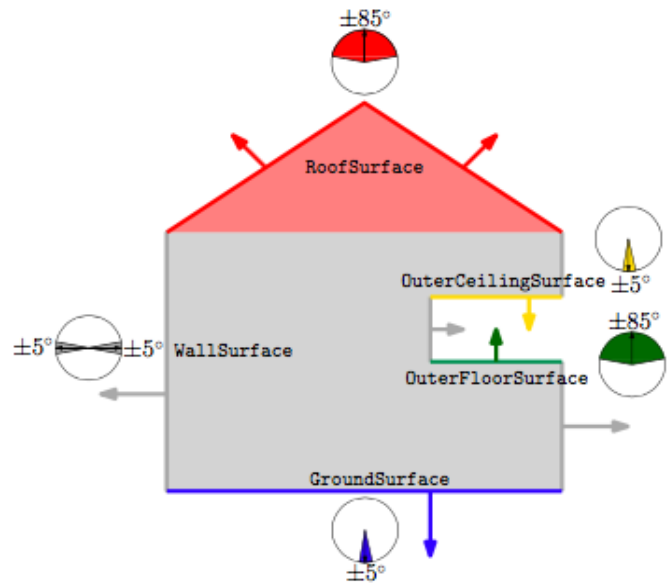

**Figure 44 - The semantic classes for buildings, and the angles and tolerances as mentioned by [12].**

# CHAPTER 5

# SUMMARY AND FUTURE WORK

## SUMMARY

The methodology presented in this dissertation and its prototype implementation help create semantically valid CityGML models from OpenSceneGraph/OBJ models. The method can be extended to CityGML classes other than buildings.

The time taken to map semantics of a building object is greatly reduced by the segmentation tools provided in the prototype. Simple building models can be semantically mapped into various classifications with just a few clicks. Added custom functions to classes have not been covered. SRS  location data have not been covered but most viewers allow these to be set when viewing the model.

Initial analysis show that LOD4 models can also be created using similar approach by extending the method and implementation to incorporate LOD4 classes. However automatic segmentation of classes have not been investigated.

# FUTURE WORK

A lot of work can be automated by the analysis of geometry such as the normal vector, the location of surfaces relative to the bounding box edges.

Validation of user assigned semantics based on a certain pre-determined range for the normal is also possible although these may not hold true for every situation, e.g. a semi-circular roof structure may have surface normal pointing a range of $\pm 180°$.

The current method can be extended to CityGML object of other types such as bridges etc. The current prototype only supports linear rings and other types of GML geometry can be added. All the surfaces are stored as triangle primitives however this can be optimised to output triangle strips and other primitives that occupy less memory.

Machine learning methods can be applied to learn features from user input to automate classifications of features and only the ambiguous classes can then be left for the user to classify. An example would be setting of weight based on the analysis of geometry from previously generated CityGML models using the prototype.